\documentclass[preprintnumbers,amsmath,amssymb,floatfix,10pt,prd,twocolumn,
superscriptaddress,nofootinbib]{revtex4}

\usepackage[parfill]{parskip}    
\usepackage{graphicx}
\usepackage{amssymb}
\usepackage{epstopdf}
\usepackage{color}
\usepackage{float}
\usepackage{amsmath}
\usepackage{orcidlink}
\usepackage{adjustbox}
\usepackage{hyperref}

\begin{document}

\title{On static and rotating decoupled black holes without inner horizons}

\author{Pablo León
 \orcidlink{0000-0002-7104-5746}}
\email{pleon@perimeterinstitute.ca}
\affiliation{Perimeter Institute for Theoretical Physics,
Waterloo, ON N2L 2Y5, Canada.}

\author{B. Mishra
\orcidlink{0000-0001-5527-3565}}
\email{bivu@hyderabad.bits-pilani.ac.in}
\affiliation{Department of Mathematics, Birla Institute of Technology and Science-Pilani, Hyderabad Campus, Hyderabad 500078, India.}

\author{Y. Gómez-Leyton}
\email{y.gomez@ucn.cl}
\affiliation{ Institute Millennium for Research on Volcanic Risk, Ckelar Volcanes, Avenida Angamos
0610, Antofagasta, Chile.}
\affiliation{Departamento de Física, Universidad Católica del Norte, Av. Angamos 0610, Antofagasta, Chile.
}

\author{Francisco Tello-Ortiz \orcidlink{0000-0002-7104-5746}}
\email{francisco.tello@ufrontera.cl}
\affiliation{Departamento de Ciencias Físicas, Universidad de La Frontera, Casilla 54-D, 4811186 Temuco, Chile.}

\begin{abstract}
Through gravitational decoupling using the extended minimal geometric deformation, a new family of static and rotating ``hairy'' black holes is provided. The background of these models is a generic Schwarzschild metric containing as special cases, the Schwarzschild, Schwarzschild-dS, Reissner-Nordstr\"{o}m and  Reissner-Nordstr\"{o}m-dS black holes. Assuming the Kerr-Schild condition and a general equation of state, the unknown matter sector is solved given rise to black hole space-times 
without a Cauchy horizon, transforming the original time-like singularity of the Reissner-Nordstr\"{o}m and Reissner-Nordstr\"{o}m-dS black holes into a space-like singularity. This fact is preserved for the rotating version of all these solutions. 
\end{abstract}
\maketitle

\section{Introduction}

The black hole (BH) physics field, constitutes one of the most impressive and active research area. The imaging of BH shadows by the Event Horizon Telescope (EHT) \cite{EventHorizonTelescope:2019dse,EventHorizonTelescope:2019uob,EventHorizonTelescope:2019jan,EventHorizonTelescope:2019ths,EventHorizonTelescope:2019pgp,EventHorizonTelescope:2019ggy,Vagnozzi:2022moj}  has ushered a new epoch, providing unprecedented opportunities to delve into the fundamental nature of BHs. 

A widely known fact is that the gravitational field of BHs are determined only by parameters such as mass and angular momentum, while the electromagnetic charges having a less impact ~\cite{Hawking:1971vc}. 
Such fact is supported by the no-hair theorem, which asserts that these solutions should not carry any other charges~\cite{Ruffini:1971bza}. Importantly, due to inner gauge symmetries, BHs can also have ``hairs'', i.e., additional charges which parameterize the BH solution \cite{Hawking:2016msc}, ``bypassing'' somehow the no-hair theorem. However, the inclusion of new parameters coming from conserved charges, also modify the BH causal structure \cite{Dadhich:2020ukj}. For example, the well-known Reissner-Nordstr\"{o}m (RN), Kerr and Kerr-Newman (KN) BHs, all of them having an event horizon and an inner horizon, the so-called Cauchy horizon. It appears once a small amount of charge is ``placed on'' the BH or once it is set to rotate. This surface is linked with the strong cosmic censorchip conjecture \cite{Penrose:1964wq,Penrose:1969pc}, requiring this conjecture  that the Cauchy horizon being unstable against perturbations in order to avoid the problem of
indeterminism. This is so because, the geometries below the
Cauchy horizon in the RN, Kerr and KN solutions are not plausible from the physical point of view. What is more, it is expected that such instabilities destroy the Cauchy horizon and the singularity cover by this surface becomes space-like \cite{Chesler:2019tco,Davey:2024xvd} (and references therein). These instabilities produce the so-called ``blueshift instability'' or ``mass inflation'', where incoming matter and radiation experience an extreme amplification in energy near the Cauchy horizon, causing a backreaction effect on space-time geometry \cite{Ong:2020xwv}.

In any case, the presence of a Cauchy horizon is rather problematic, so it is desirable to avoid it, and this work focuses on this issue in the framework of gravitational decoupling (GD) \cite{Ovalle:2017fgl,Ovalle:2018gic,Ovalle:2020fuo}. Therefore, the main goal of this article, is to construct BH solutions employing GD\footnote{For recent developments of BH physics in the context of gravitational decoupling, see \cite{Heydarzade:2023dof,Ovalle:2018umz,Cavalcanti:2022adb,Cavalcanti:2022cga,Meert:2021khi,Sultana:2021cvq,Ovalle:2020kpd,daRocha:2020gee,Fernandes-Silva:2019fez,Casadio:2022ndh,Ovalle:2022eqb,Estrada:2020ptc,Estrada:2021kuj,Ovalle:2023ref,Zhang:2022niv,Khosravipoor:2023jsl,Casadio:2024uwj,Liang:2024xif,Ovalle:2021jzf,Contreras:2021yxe}.}, so that the salient model being free of inner horizons. To derive the solutions for BH scenarios, we can pursue two strategies. The simplest approach involves considering minimal modifications to the geometry using the Minimal Geometric Deformation (MGD) method \cite{Ovalle:2017fgl}. In this case, the generic matter sector solely interacts gravitationally with the original fluid that supports the undeformed solution which ensures that energy-momentum conservation is maintained for each fluid separately. The alternative strategy involves completely deforming the metric, the extended version of MGD (e-MGD) \cite{Ovalle:2018gic}, guaranteeing a well-defined killing horizon. In this work, we shall explore the second approach because it is possible to keep the usual Schwarzschild BH form, namely $g_{tt}(r)g_{rr}(r)=-1$, which is essential in achieving BHs without Cauchy horizons. Besides, this fact allows us to reduce the number of degrees of freedom and also prevent nonphysical signature changes. Additionally, we supply the generic unknown matter sector  with a general equation of state (EoS) to close the system. Interestingly, these constraints lead to a Schwarzschild-like BH with a polynomial correction. Under certain conditions, this correction, characterized by a ``secondary hair'' changes the causal structure of the BH region, delinting the inner horizon. To go beyond the static scenario, we extend our study to a more realistic arena obtaining the rotating version and studying some relevant properties such as the BH shadow. It is worth mentioning that the staring point or seed space-time (background) of these hairy models, is a family of generic Schwarzschild BH, all of them solutions of the General Relativity (GR) theory, containing the vacuum Schwarzschild, Schwarzschild-dS, RN and RN-dS BHs.

This work is organized as follows. The next section is devoted to introduce the main aspects of GD. Next, in section \ref{section3}, we construct the static BH solutions by implementing a general EoS. In section \ref{section4}, we obtain and analyze the rotating version of the static solutions, studying some of their properties and  section \ref{section5} concludes the work.

 Throughout the article we use the mostly negative signature $\{+;-;-;-\}$ and units where $c=G_N=1$, thus $\kappa=8\pi$.

\section{Gravitational decoupling}\label{section2}
In this section, we briefly describe the so--called GD in its extended fashion, that is, e--MGD for spherically symmetric space-times, described in details \cite{Ovalle:2018gic}. This case represents a more general setup than the MGD case \cite{Ovalle:2017fgl}. We consider the Einstein field equations 

\begin{align}
\label{EinEqFull}
G_{\mu\nu}\equiv R_{\mu \nu} - \frac{1}{2}g_{\mu \nu} R = \kappa {T}_{\mu \nu},
\end{align}
with a total energy-momentum tensor containing two contributions,
\begin{equation}
T_{\mu\nu}=\tilde{T}_{\mu\nu}+\theta_{\mu\nu},
\end{equation}
where $\tilde{T}_{\mu\nu}$ is usually associated with some known solution
of general relativity, whereas $\theta_{\mu\nu}$ may contain new fields or a new gravitational sector.

The equations of motion (\ref{EinEqFull}) along with a generic spherically symmetric and static geometry is,
\begin{equation}
ds^{2}=e^{\nu(r)}\,dt^{2}-e^{\lambda(r) }\,dr^{2}
-r^{2}d\Omega^{2}, \label{metric}
\end{equation}
where $d\Omega$ is the usual two sphere angular part, provide the following system of equations 
\begin{align}
\label{ec1}
\kappa T^{0}_{0}(r) &= \frac{1}{r^2}-e^{-\lambda(r)}\left[\frac{1}{r^2}-\frac{\lambda'(r)}{r}\right]\,,
\\\label{ec2}
\kappa T^{1}_{1}(r) &= \frac{1}{r^2}-e^{-\lambda(r)}\left[\frac{1}{r^2}+\frac{\nu'(r)}{r}\right]\,,
\\\label{ec3}
\kappa T^{2}_{2}(r) &=-\frac{1}{4}e^{-\lambda(r)}\bigg[2 \nu ''(r) + \nu'^{2}(r)-\lambda ' (r)\nu'(r)  \nonumber\\
&+ 2 \frac{\nu'(r)-\lambda'(r)}{r}\bigg],
\end{align}
where primes denote differentiation with respect to the radial coordinate $r$ and $T^{2}_{2}=T^{3}_{3}$. From (\ref{ec1})--(\ref{ec3}) one can recognize the following effective quantities,
\begin{align}
\label{den}
\rho(r)\equiv T^{0}_{0}(r)&=\tilde{T}^{0}_{0}(r)+\theta^{0}_{0}(r),\\ \label{prerad}  p_{r}(r)\equiv  T^{1}_{1}(r)&=-\tilde{T}^{1}_{1}(r)-\theta^{1}_{1}(r), \\ \label{tanpre}
p_{\perp}(r)\equiv  T^{2}_{2}(r)&=-\tilde{T}^{2}_{2}(r)-\theta^{2}_{2}(r),
\end{align}
that is, an effective density, radial and tangential pressures. 

At this stage, it should be pointed out that as the Einstein's tensor $G_{\mu\nu}$ satisfies Bianchi identity, $\nabla_{\mu}G^{\mu\nu}=0$, so Eq. (\ref{EinEqFull}) becomes,
\begin{equation}\label{tmunutotal}
\nabla_{\mu}T^{\mu\nu}=0 \Rightarrow \nabla_{\mu}\tilde{T}^{\mu\nu}+\nabla_{\mu}\theta^{\mu\nu}=0.
\end{equation}
Nevertheless, the above conservation law is subjected to the following conditions, namely i) both energy--momentum tensors, $\tilde{T}^{\mu\nu}$ and $\theta^{\mu\nu}$, are separately conserved,  ii) $\nabla_{\mu}\tilde{T}^{\mu\nu}=-\nabla_{\mu}\theta^{\mu\nu}$. The former means that the gravitational sources only interact gravitationally \cite{Ovalle:2017fgl}, while the second case says that there is an energy exchange between the sources \cite{Ovalle:2022yjl}. 

Now, the solution of the system (\ref{ec1})--(\ref{ec3}) requires the input of the so--called seed spacetime
\begin{equation}\label{seedsp}
d s^2=e^{\xi(r)} d t^2-e^{\mu(r)} d r^2-r^{2}d\Omega^{2},
\end{equation}
characterized by the seed energy--momentum tensor $\tilde{T}_{\mu\nu}$. The incorporation of the $\theta$--sector, allows us to introduce the following deformation functions $g(r)$ and $f(r)$
\begin{eqnarray}\label{expectg}
\xi(r) &\to& \nu(r)=\xi(r)+ \alpha g(r), \\ \label{expectg1}
e^{-\mu(r)}&\to&e^{-\lambda(r)}=e^{-\mu(r)}+\alpha f(r),
\end{eqnarray}
in order to decouple the source $\tilde{T}_{\mu\nu}$ from the source $\theta_{\mu\nu}$, in such a way that the intricate set of Eqs. (\ref{ec1})--(\ref{ec3}), splits into two separate systems. To achieve this, the e--MGD (\ref{expectg})--(\ref{expectg1}) should be replaced into the set (\ref{ec1})--(\ref{ec3}), leading to 
\begin{align} \label{ro1}
\kappa \tilde{\rho}(r)&=\frac{1}{r^2}-e^{-\mu(r)}\left[\frac{1}{r^2}-\frac{\mu^{\prime}(r)}{r}\right], \\ \label{pr1}
 \kappa \tilde{p}_r(r)&=-\frac{1}{r^2}+e^{-\mu(r)}\left[\frac{1}{r^2}+\frac{\xi^{\prime}(r)}{r}\right], \\ \label{pt1}
 \kappa \tilde{p}_{\perp}(r)&=\frac{e^{-\mu(r)}}{4}\bigg[2 \xi^{\prime \prime}(r)+\xi^{\prime 2}(r)-\mu^{\prime}(r) \xi^{\prime}(r)\\ & \nonumber
 +2 \frac{\xi^{\prime}(r)-\mu^{\prime}(r)}{r}\bigg],
\end{align}
corresponding to the usual Einstein's field equations and the second one is a system sourced by the $\theta$--sector given by
\begin{align}
 \label{ec1d}
\kappa \theta^{0}_{0}(r)&=-\frac{\alpha f(r)}{r^2}-\frac{\alpha f'(r)}{r}, \\ \label{ec2d}
 \kappa \theta^{1}_{1}(r)+\alpha Z_1(r)&=-\alpha f(r)\left[\frac{1}{r^2}+\frac{\nu'(r)}{r}\right], \\\nonumber
\kappa \theta^{2}_{2}+\alpha Z_2(r)&=-\frac{\alpha f(r)}{4}\left[2 \nu''(r)+{\nu'(r)}^{2}+2 \frac{\nu'(r)}{r}\right] \\ & \label{ec3d} -\frac{\alpha f'(r)}{4}\left[\nu'(r)+\frac{2}{r}\right], 
\end{align}
where $Z_{1}(r)$ and $Z_{2}(r)$ are defined as
\begin{align}
Z_1(r) & =\frac{e^{-\mu(r)} g'(r)}{r}, \\
4 Z_2(r) & =e^{-\mu(r)}\bigg[2 g''(r)+\alpha {g'}^{2}(r)+\frac{2 g'(r)}{r}+2 \xi'(r) g'(r) \nonumber \\
&-\mu'(r) g'(r)\bigg].
\end{align}
From the above set of equations, it is evident that the $\theta_{\mu\nu}$ source vanishes when both, $f(r)$ and $g(r)$ are zero. As pointed out before, the conservation of the full energy--momentum tensor $T_{\mu\nu}$ is guaranteed if and only if each sector, the seed and the new one, are separately conserved or under an exchange of energy between them. Particularly, the Eq. (\ref{tmunutotal}) provides
\begin{equation}\label{conservation1}
\nabla_\mu \tilde{T}_\nu^\mu=-\frac{\alpha g'(r)}{2}\left[\tilde{\rho}(r)+\tilde{p}_{r}(r)\right] \delta^{r}_{\nu}=-\nabla_\mu \theta_\nu^\mu.
\end{equation}
This shows an exchange of energy between the sources $\tilde{T}_{\mu\nu}$ and $\theta_{\mu\nu}$, where the temporal deformation function $g(r)$ plays a major role. Therefore, a pure gravitational interaction comes from taking $g(r)=0$, that is, on the so--called MGD case. Another way to obtain a gravitational interaction only, is by considering the seed spacetime belonging to the class of the Kerr--Schild spacetimes \cite{Kerr:1965wfc}. In such a case, one has $g_{tt}(r)g_{rr}(r)=-1$ implying ${p}_{r}(r)=-{\rho}(r)$. 

\section{Black holes}\label{section3}

The approach we follow here to find  BHs without inner horizon, starting from spherically
symmetric BHs in general relativity, is based on fully deforming the well-known generic Schwarzschild metric
\begin{equation}\label{schw}
    e^{\xi(r)}=e^{-\mu(r)}=1-\frac{2M(r)}{r},
\end{equation}
which solves Eqs. (\ref{ro1})-(\ref{pt1}), being this space-time our seed geometry. It should be pointed out that, instead of taking a particular mass function $M(r)$, here we express it as a family containing  well-known BHs, namely:

\begin{enumerate}
    \item $M(r)=M$: Schwarzschild BH, subjected to $\tilde{T}_{\mu\nu}=0$.
    \item $M(r)=M+\frac{\Lambda}{6}r^{3}$: Schwarzschild-dS BH, with $\tilde{T}_{\mu\nu}=\text{diag}\{\frac{\Lambda}{8\pi}, -\frac{\Lambda}{8\pi}, -\frac{\Lambda}{8\pi}, -\frac{\Lambda}{8\pi}\}$.
    \item $M(r)=M-\frac{Q^{2}}{2r}$: Reissner-Nordstr\"{o}m BH, described by $\tilde{T}_{\mu\nu}=\text{diag}\{\frac{E^{2}}{8\pi}, -\frac{E^{2}}{8\pi},\frac{E^{2}}{8\pi} ,\frac{E^{2}}{8\pi} \}$.
    \item $M(r)=M-\frac{Q^{2}}{2r}+\frac{\Lambda}{6}r^{3}$: Reissner-Nordstr\"{o}m-dS BH, characterized by $\tilde{T}_{\mu\nu}=\text{diag}\{\frac{E^{2}}{8\pi}+\frac{\Lambda}{8\pi}, -\frac{E^{2}}{8\pi}-\frac{\Lambda}{8\pi},\frac{E^{2}}{8\pi} -\frac{\Lambda}{8\pi},\frac{E^{2}}{8\pi} -\frac{\Lambda}{8\pi}\}$.
\end{enumerate}

The main point in searching the energy-momentum tensor $\theta_{\mu\nu}$ which induces both, $g(r)$ and $f(r)$, is to remove the Cauchy horizon of the seed metric. 
Of course, the seed solutions 1. and 3. have not Cauchy horizon. In such cases, we expect BHs with simple causal structure once the $\theta$-sector is incorporated. Furthermore, for these models, in passing from the static to the rotating situation it is expect that the Cauchy horizon produced by rotating effects being suppressed by the new elements.  

Given that the $\theta$-sector system (\ref{ec1d})-(\ref{ec3d}) has five unknowns, namely $\{g(r);f(r);\theta^{0}_{0}(r);\theta^{1}_{1}(r);\theta^{2}_{2}(r)\}$, then we are free to impose additional conditions. This will be a matter of fact of the next subsections.

\subsection{Horizon structure}

The causal structure of BHs is a key feature. Therefore, to avoid any inconsistency we are going to impose the so-called Schwarzschild condition where the following relation between the metric potentials holds
\begin{equation}\label{metrice}
    e^{\nu(r)}=e^{-\lambda(r)}.
\end{equation}
In this way, the causal horizon $r=r_{H}$ leads to 
\begin{equation}\label{feos}
    e^{\nu(r)}\bigg{|_{r=r_{H}}}=e^{-\lambda(r)}\bigg{|_{r=r_{H}}}=0.
\end{equation}
Here, the causal horizon $r_{H}$ is also a Killing horizon $r_{K}$. It is worth mentioning that the above condition is not mandatory to recognize a space-time as a BH one, however, this constraint alleviates the mathematical treatment of the field equations and avoids undesirable signature changes drifting apart from the Lorentzian requirement. Additionally, from the Eqs. (\ref{ec1})-(\ref{ec3}) it is not hard to see that the condition (\ref{metrice}) produces the following EoS
\begin{equation}
 \rho(r)+p_{r}(r)=0\Rightarrow   \theta^{1}_{1}(r)=\theta^{0}_{0}(r).
\end{equation}
This result comes from the fact that $\tilde{\rho}+\tilde{p}_{r}=0$, for all the seed space-times considered here. In this way, one can have the following relation between the decoupler functions $g(r)$ and $f(r)$
\begin{equation}\label{rela}
    f(r)=e^{\nu(r)}\left(c+\int e^{-\nu(r)-\mu(r)}g'(r)dr\right),
\end{equation}
recalling that $\nu(r)$ is given by (\ref{expectg}). Particularly, for Schwarzschild-like metric potentials (\ref{schw}), the above expression takes the simple form
\begin{equation}
 \alpha f(r)=\left(1-\frac{2M(r)}{r}\right)\left(e^{\alpha g(r)}-1\right).   
\end{equation}
Therefore, the general metric (\ref{metric}) becomes
\begin{align}\label{mm}
    ds^{2}=\left(1-\frac{2M(r)}{r}\right)e^{\alpha g(r)}dt^{2}-\left(1-\frac{2M(r)}{r}\right)^{-1}\nonumber &\\ \times e^{-\alpha g(r)}dr^{2}-r^{2}d\Omega^{2},
\end{align}
where $g(r)$ is yet to be determined.

\subsection{General equation of state}

The next step consists in imposing an extra constraint
to solve for the function $g(r)$. In this concern, we are going to enforce the $\theta$-sector components to satisfy a general EoS of the form \cite{Ovalle:2018umz,Ovalle:2020kpd},
\begin{equation}\label{eosgene}
    \theta^{0}_{0}(r)=a\theta^{1}_{1}(r)+b\theta^{2}_{2}(r),
\end{equation}
being $a$ and $b$ arbitrary parameter with units of square length. Plugging (\ref{ec1d}), (\ref{ec2d}) and (\ref{ec3d}) into (\ref{eosgene}) one gets the the following differential equation
\begin{equation}\label{diff1}
    A(r)U''(r)+B(r)U'(r)+C(r)U(r)=0,
\end{equation}
where the coefficients are given by
\begin{align}
    A(r)=& br\left[r-2M(r)\right], \\ 
    B(r)=& 4\left[1-a\right]M(r)-2r\bigg[1+2bM'(r) \nonumber \\
    & -a-b\bigg], \\
    C(r)=& \left[1-a\right]\left[4M'(r)-2\right]-2brM''(r),
\end{align}
and $U(r)\equiv e^{\alpha g(r)}$ is an auxiliary function. Now, the expression (\ref{diff1}) can be formally integrated and providing 
\begin{align}\label{uu}
    U(r)=\frac{1}{\left[r-2M(r)\right]}\bigg[c_{1}-\frac{1}{D}\int C(r)dr   &\\ \nonumber+\frac{r^{D/b}}{D}\left(c_{2}+\int r^{-D/b}\,C(r)dr\right)\bigg],
\end{align}
where $c_{1}$ and $c_{2}$ are integration constants with units of length and $\text{length}^{3-D/b}$, respectively, and $D\equiv 2-2a+b$.

\subsection{Hairy black holes}

Using the expressions (\ref{mm}) and  (\ref{uu}), we obtain the following metric
\begin{align}\label{result1}
  e^{\nu(r)}=e^{-\lambda(r)}=\frac{1}{r}\bigg[c_{1}-\frac{1}{D}\int C(r)dr   &\\ \nonumber+\frac{r^{D/b}}{D}\left(c_{2}+\int r^{-D/b}\,C(r)dr\right)\bigg].
\end{align}
Now, instead to express every separate case for the seed mass function $M(r)$, we are to consider the most general one, that is, the Reissner-Nordstr\"{o}m-dS case and the remaining cases can be analyzed by taking the appropriate limit. Therefore, one can obtain
\begin{equation}
     e^{\nu(r)}=e^{-\lambda(r)}=1-\frac{c_{1}}{r}+\frac{Q^{2}}{r^{2}}-\frac{\Lambda}{3}r^{2}+\frac{l}{r^{n}},
\end{equation}
where the parameter $l$ is a combination of the previous constants and can be considered in principle as primary hair. On the other hand, the seed mass parameter $M$ does not appear in the
solution, then the mass is instead given by $\mathcal{M}\equiv c_{1}/2$. In this way one gets
\begin{equation}\label{fiso}
     e^{\nu(r)}=e^{-\lambda(r)}=1-\frac{2\mathcal{M}}{r}+\frac{Q^{2}}{r^{2}}-\frac{\Lambda}{3}r^{2}+\frac{l}{r^{n}},
\end{equation}
besides, $n\equiv-\frac{2}{b}(a-1)$, where we are going to consider for the sake of simplicity, only those values corresponding to $n\in \mathbb{Z}^{+}$. To get non-trivial BHs, we impose the additional restriction $n\geq 3$, because cases $n=1,2$ are the Schwarzschild and RN BHs up to a redefinition of the mass and charge parameters.
As can be seen from (\ref{fiso}), the asymptotic behavior of the seed space-time is preserved. This is so because at large enough distances the correction $1/r^{n}$ falls off faster than the other terms.

Now, possible horizons are found from solutions $r_{H}=r_{H}(\mathcal{M},Q,\Lambda,l)$ of
\begin{equation}
    e^{-\lambda(r_{H})}=0.
\end{equation}
As we are interested in deleting the original Cauchy horizon of the RN and RN-dS models ({and also avoiding the introduction of additional internal causal structure for the Schwarzschild and Schwarzschild-dS cases}) taking advantage of the new term, we need to carefully analyze under what conditions the $(n+2)$-degree polynomial expression (\ref{fiso}) provides more than one positive real root\footnote{ Particularly in this work, we are going to consider the dS case. The AdS case can be done in the same way.}. So, reorganizing the polynomial expression one gets
\begin{equation}\label{eq40}
    -\frac{\Lambda}{3}r_{H}^{n+2}+r_{H}^{n}-2\mathcal{M}r_{H}^{n-1}+Q^{2}r_{H}^{n-2}+l=0. 
\end{equation}
Next, one needs to take into account the following cases: i) Odd $n$ and ii) even $n$, both subject to the condition $l>0$ and $l<0$. Let's start by considering the case $l<0$, then the above polynomial can be recast as 
\begin{equation}\label{case1}
    -\frac{\Lambda}{3}r_{H}^{n+2}+r_{H}^{n}-2\mathcal{M}r_{H}^{n-1}+Q^{2}r_{H}^{n-2}-|l|=0. 
\end{equation}
It is not hard in general to show that for odd $n$, the expression (\ref{case1}) has a maximum of four positive real roots and a minimum of two. The rest of the roots correspond to negative and complex values. Obviously, negative and complex roots have no physical meaning and lead to naked singularities. The even $n$ case is similar, that is, a maximum of four real positive roots and a minimum of two real positive roots. So, it is clear that when $l$ is negative in nature, the BH causal structure is modified, that is, the BH has acquires one more inner horizon, keeping its event and cosmological horizons or there is no longer an inner horizon. Therefore, the case $l<0$ is of interest here, because we are looking for those BH solutions with a simple internal causal structure, without inner horizons.  
Now, when $l>0$ one has 
\begin{equation}\label{case2}
    -\frac{\Lambda}{3}r_{H}^{n+2}+r_{H}^{n}-2\mathcal{M}r_{H}^{n-1}+Q^{2}r_{H}^{n-2}+l=0. 
\end{equation}

\begin{figure}[h]
\centering
\includegraphics[width=0.4\textwidth]{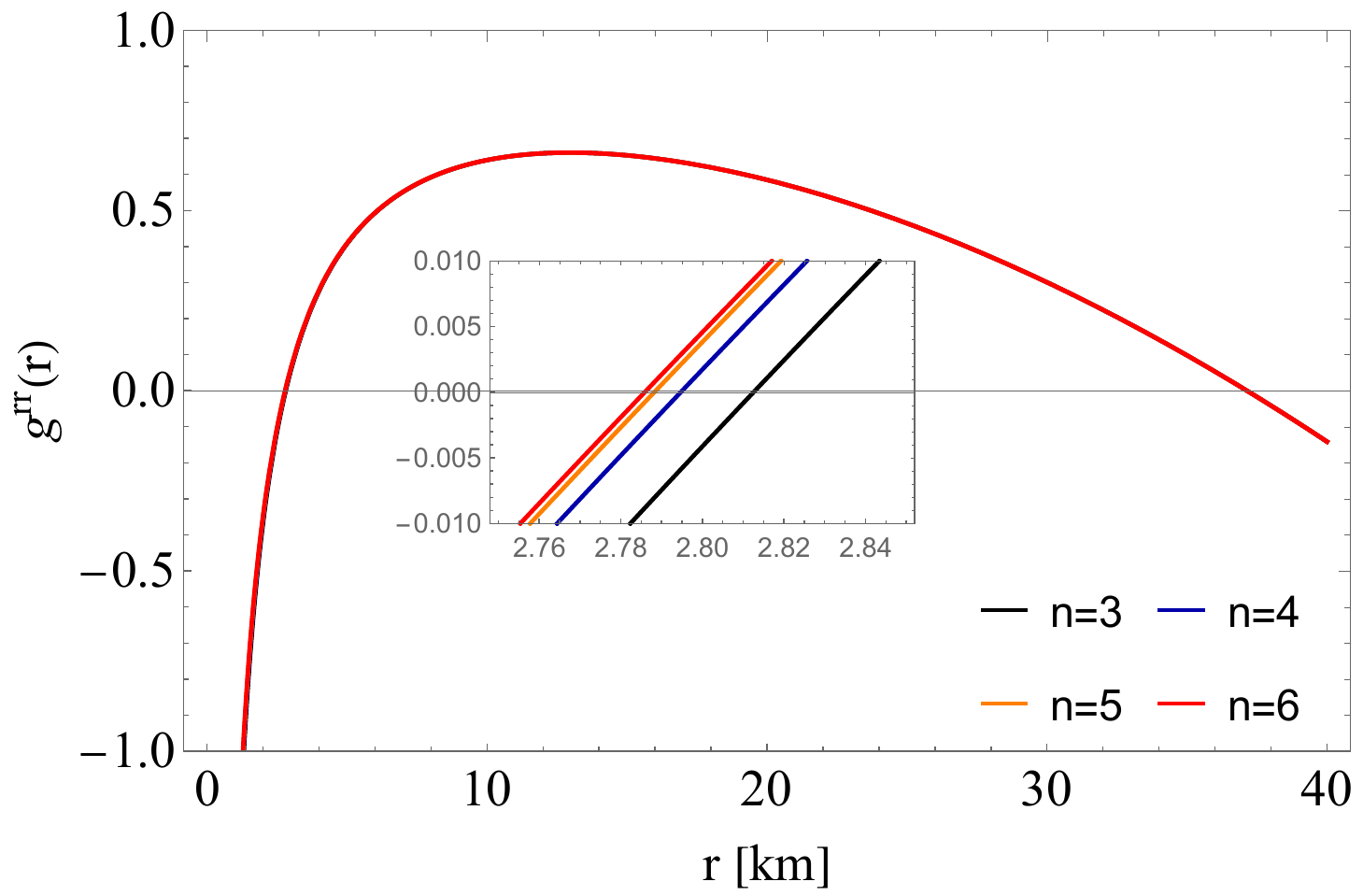} \
\includegraphics[width=0.4\textwidth]{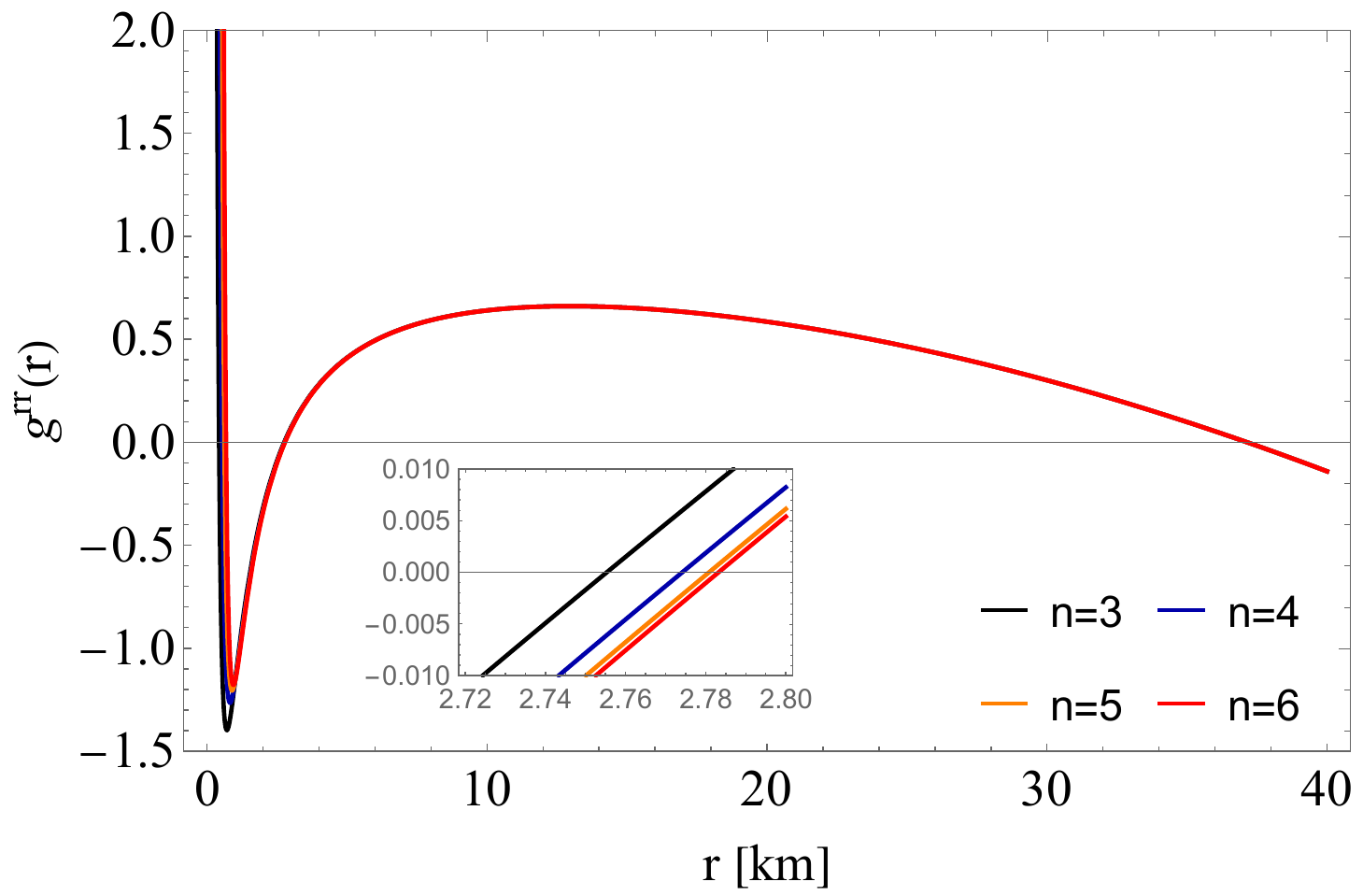} 
\caption{ \textbf{Upper panel}: The trend of the inverse radial metric potential versus the radial coordinate. This shows case for $l<0$, exhibiting only the event and the cosmological horizons. \textbf{Lower panel}: Again, the behavior of the inverse radial metric potential against the radial coordinate. Here we consider $l>0$, then the solutions besides the event and cosmological horizons, presents a Cauchy horizon. For these panels we consider $\mathcal{M}=1.5$ [km], $Q=0.8$ [km], $\Lambda=0.002$ [$\text{km}^{-2}$] and $l=\pm0.2$ [$\text{km}^{n}$]. \label{fig1}}
\end{figure}
From here one obtains for both, odd and even $n$, a BH space-time with same causal structure as the seed solution, that it, a BH with a Cauchy, event and cosmological horizons. However, there is also a possibility of having a naked singularity. In conclusion, the case $l>0$ keeps the original causal structure independent of the inclusion of the parameter $l$ or leads to a naked singularity protected at most for the cosmological horizon.  

So, to assure only two positive distinct real roots, we need to demand that the discriminant of the polynomial (\ref{case1}) must be positive in nature, that is
\begin{equation}
D(P(r_{H}))=(-1)^{n(n-1) / 2} a_n^{2 n-2} \prod_{i \neq j}\left(r_{Hi}-r_{Hj}\right)>0,
\end{equation}
where in this case $a_{n}=-\Lambda/3$, with $\Lambda>0$.

{Albeit the case $l<0$ leads to a BH with a simple causal structure, we can not assure at all that $l\in (-\infty,0)$ is always valid. This is so because it may be happen that for certain values of the parameter $l$ the BH region is containing more than one inner horizon (a degenerated Cauchy horizon). Furthermore, as we are dealing with a polynomial of degree $n+2$, it is not an easy task to restrict or bound the possible values that $l$ could take to avoid a multi-horizon BH\footnote{One way to bound the magnitude of the parameter $l$ in terms of the mass $\mathcal{M}$, charge $Q$ and cosmological constant $\Lambda$ is through the discriminant of the polynomial. However, this is not possible here due to the degree of the polynomial expression. }. Nevertheless, analyzing the behavior of the strong energy condition (SEC) at the BH region, we can get some insights about the range of the parameter $l$. As it is well-known, the number of inner horizons can be restricted imposing the violation of the SEC \cite{Yang:2021civ} at some point satisfying $0<\tilde{r}<r_{H}$. The SEC states that \cite{Curiel:2014zba}
\begin{eqnarray}\label{43}
   \rho(r)+p_{i}(r)&\geq& 0, \\ \label{45}
      \rho(r)+p_{r}(r)+2p_{\perp}(r)&\geq& 0,
\end{eqnarray}     
with $i=r,\perp$. As (\ref{43}) is fulfilled in the radial direction, inserting this result into (\ref{45}) one obtains 
\begin{equation}
    p_{\perp}\geq 0.
\end{equation}
Therefore, to violate SEC in the BH region is it necessary just to impose 
\begin{equation}\label{const}
    p_{\perp}<0 \Rightarrow l<\frac{2\tilde{r}^{\,n-2}\left(\tilde{r}^{4}\Lambda-Q^{2}\right)}{\left(n-1\right)n}.
\end{equation}
As can be observed, this result is compatible with the assumption $l<0$. Notwithstanding, as observed, $l$ can not be any negative number. Of course, the size of $\tilde{r}$ is intimately related with the size of the event horizon and this one is determined by the values of the parameter space $\{\mathcal{M}, Q, \Lambda, l\}$. In Fig. \ref{fig1} it is displayed the trend of the radial metric potential for $l<0$ (upper panel) and $l>0$ (lower panel). For the chosen values of the parameter space and taking into account the constraint (\ref{const}), the inner horizon has been avoided for the case $l<0$ and no additional structure has been included for the case $l>0$, that is, there are only the Cauchy, event and cosmological horizons. As can be observed, in absent of Cauchy horizon, the even horizon increases its size as $n$ decreases, contrary to what happens when the BH has an inner horizon, in that case the the event horizon size increases as $n$ takes greater values.  
It is worth mentioning that a multi-horizon BH (case $l<0$ for some values of the parameter $l$) is possible since this fact is associated with the introduction of some new ``charges'' \cite{Dadhich:2020ukj}. Although $l$ can be considered as genuine hair, it is not related with any global charge, namely the mass, the 
(Maxwellian) electric charge or the angular momentum. Nevertheless, the key aspect in averting the inner horizon is a twofold fact: i) This can potentially prevent the extreme instabilities associated with mass inflation and maintain the integrity of the event horizon structure and ii) the central singularity of the BH cannot be a time-like one, it is always  space-like in nature In this way, we can say that the $\theta$-sector plays a pivotal role in determining the causal structure of full deformed BHs, either by radically changing the structure of space-time or maintaining the original structure.}

\begin{figure}[H]
\centering
\includegraphics[width=0.45\textwidth]{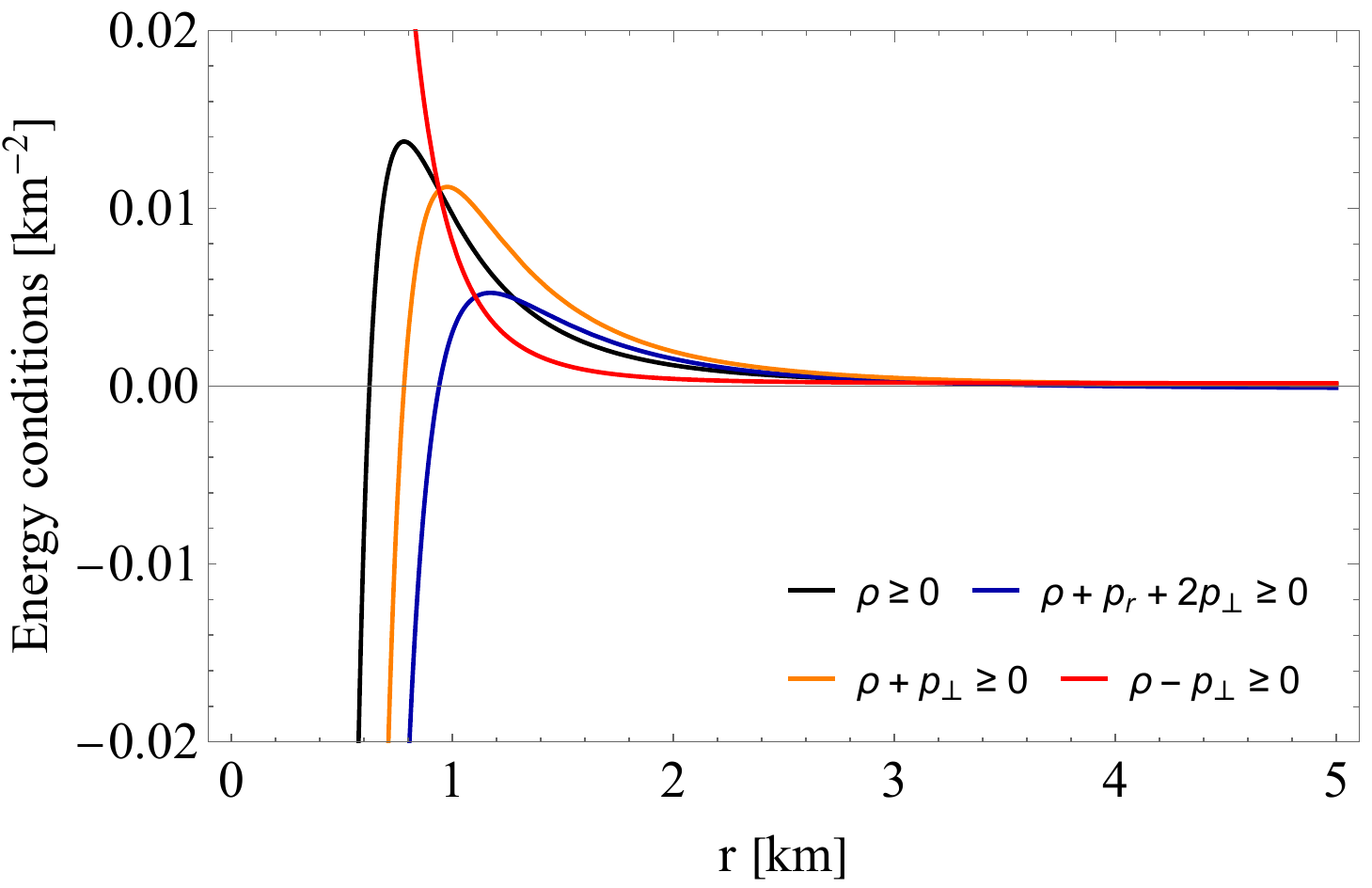}
\caption{ The trend of some inequalities corresponding to the strong and dominant energy conditions against the radial coordinate. For this plot we used $\mathcal{M}=1.5$ [km], $Q=0.8$ [km], $\Lambda=0.002$ [$\text{km}^{-2}$], $l=-0.2$ [$\text{km}^{3}$], $n=3$ and $a=0.9$. \label{fig4}}
\end{figure}

{Another important energy condition is the so-called  dominant energy condition (DEC), this says \cite{Curiel:2014zba} 
\begin{eqnarray}\label{rhodec}
\epsilon(r)&\geq& 0, \\
\label{431}
    \epsilon(r)+p_{i}(r)&\geq& 0, \\ \label{441}
     \epsilon(r)-p_{i}(r)&\geq& 0,
\end{eqnarray}
where $i=r,\perp$. The satisfaction of DEC, is related with the preservation of the event horizon topology. So, violation of this condition could induce topology changes on the event horizon. Here, the inequality (\ref{441}) in the radial direction, is just the condition (\ref{rhodec}), and as we are dealing with an asymptotically dS space-time, this means that at from the event horizon at to large enough distances (cosmological horizon) the cosmological constant dominates. Therefore, in this regime the energy density of the whole space-time will be positive in nature, satisfying (\ref{rhodec}) and both (\ref{431}) and (\ref{441}) in the radial direction. Interestingly, inequalities (\ref{43}) in the angular direction (consequently (\ref{431})), is respected when (\ref{const}) is taken into account. }

\section{Axially symmetric case}\label{section4}

Here we are going to obtain the rotating version of the toy model given by (\ref{fiso}) following the strategy described in \cite{Ovalle:2021jzf,Contreras:2021yxe}. This simply amounts to consider
the general Kerr-Schild metric in Boyer-Lindquist
coordinates, namely, the Gurses-Gursey metric

\begin{align}\label{gg}
d s^2= & \left[1-\frac{2 r \tilde{m}(r)}{\varrho^2}\right] d t^2+\frac{4 a r \tilde{m}(r) \sin ^2 \theta}{\varrho^2} d t d \phi \\ \nonumber
& -\frac{\varrho^2}{\Delta} d r^2-\varrho^2 d \theta^2-\frac{\Sigma \sin ^2 \theta}{\rho^2} d \phi^2,
\end{align}
with
\begin{equation}
\begin{aligned}
\varrho^2 & =r^2+a^2 \cos ^2 \theta \\
\Delta & =r^2-2 r \tilde{m}(r)+a^2 \\
\Sigma & =\left(r^2+a^2\right)^2-\Delta a^2 \sin ^2 \theta \\
a & =J / \mathcal{M},
\end{aligned}
\end{equation}
where $\tilde{m}(r)$ is the mass function of our reference spherically
symmetric metric (\ref{fiso}) given by
\begin{equation}
    \tilde{m}(r)=\mathcal{M}-\frac{Q^{2}}{2r}-\frac{|\Lambda|}{6}r^{3}-\frac{l}{2r^{n-1}},
\end{equation}
where classical solutions are recovered in the limit $l\rightarrow0$. Now, The line-element (\ref{gg}) contains two potential singularities, namely, when $\varrho=0$ or $\Delta=0$. The case $\varrho=0$ is the ring singularity, and it is a physical singularity which occurs at $\theta=\pi/2$ and $r=0$. As usual, the region $\Delta=0$ represents a coordinate singularity
that indicates the existence of horizons, defined
by
\begin{equation}\label{plo}
\Delta\left(r_{{H}}\right)=r_{{H}}^2-2 r_{{H}} \tilde{m}\left(r_{{H}}\right)+a^2=0.
\end{equation} 
Explicitly this polynomial reads
\begin{equation}\label{polyrot}
    -\frac{\Lambda}{3}r^{n+4}_{H}+r^{n+2}_{H}-2\mathcal{M}r^{n+1}_{H}+(Q^{2}+a^{2})r^{n}_{H}+lr^{2}=0.
\end{equation}
{As can be appreciated, the above expression can be factor out by $r^{2}$. Then, one recast almost the same expression given for the static case (\ref{eq40}). The main different is of course the presence of the term $a$. }
In this case, the above polynomial expression is quite involved leading to a more intricate causal structure. As occurs in the static case, where the Maxwellian charge induces a Cauchy horizon (RN and RN-dS BHs), when rotation is included also an inner horizon appears, such as in the Kerr BH. The main conclusion about this fact, is that matter distributions modify the causal structure of the BH region, resulting in an opposed  repulsive contribution to mass.
Curiously, when charge and rotation are present (KN solution), both the charge and angular momentum combine their effects to produce only one inner horizon. Taking into account this fact, 
in principle, we expect that the parameter $l$ is not acting in an independent way with respect to the other parameters. 

\begin{figure}[h]
\centering
\includegraphics[width=0.4\textwidth]{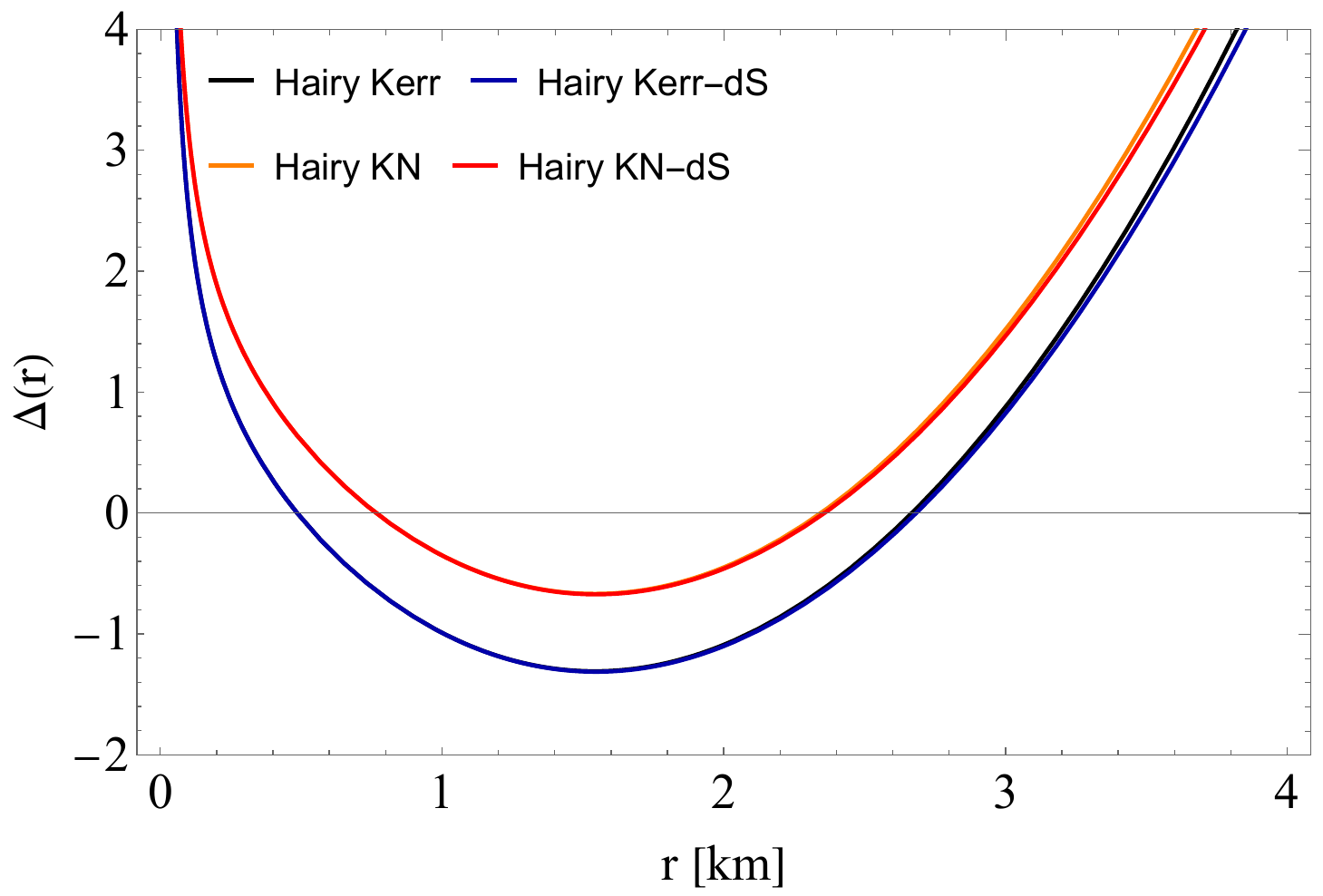} \
\includegraphics[width=0.4\textwidth]{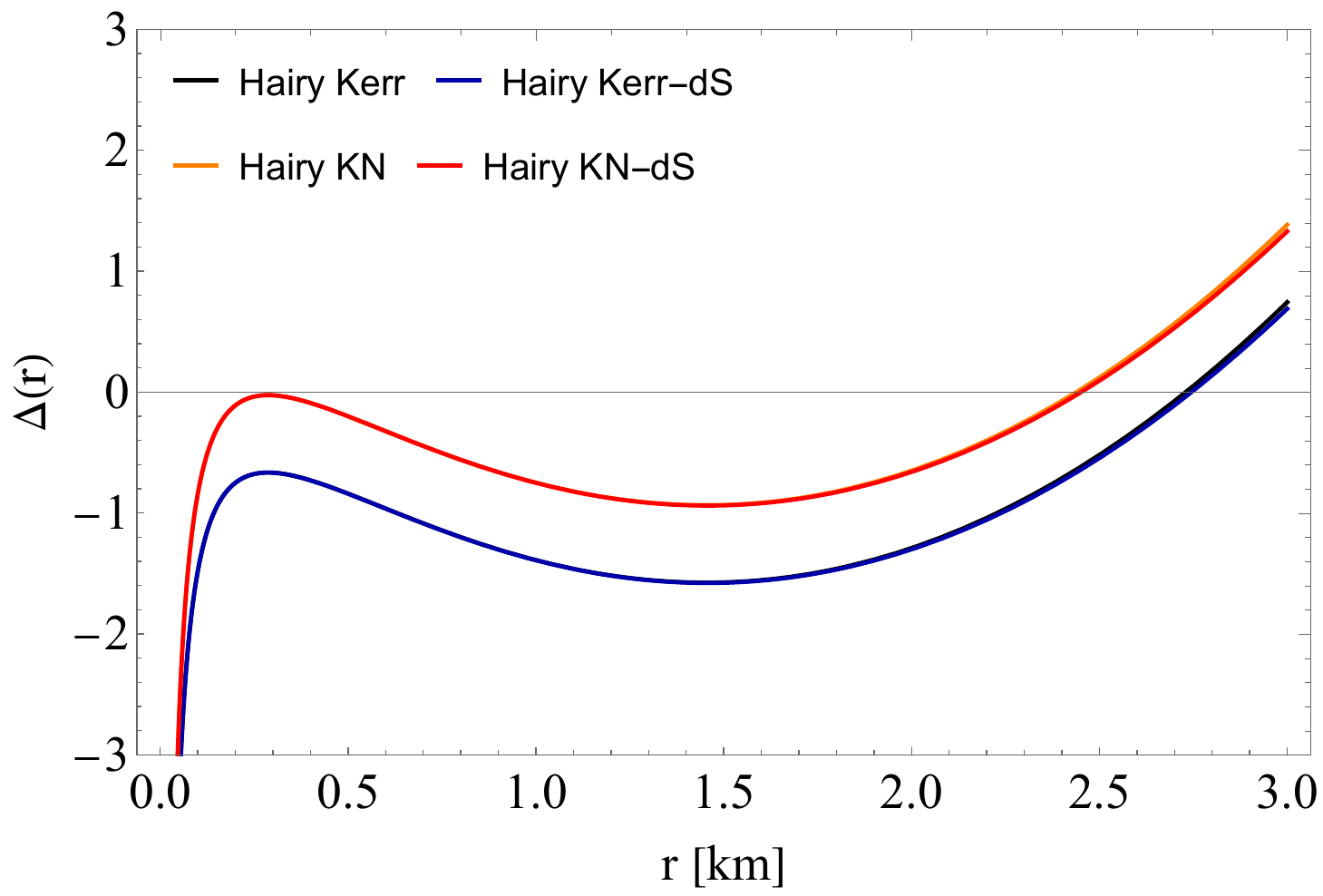} 
\caption{\textbf{Upper panel}: The trend of the function $\Delta(r)$ versus the radial coordinate $r$ for case $l>0$. \textbf{Lower panel}: The behavior of the function $\Delta(r)$ against the radial coordinate $r$ for the ``hairy'' BH solutions when $l<0$. For all these plots we used $\mathcal{M}=1.5$ [km], $Q=0.8$ [km], $\Lambda=0.002$ [$\text{km}^{-2}$], $l=\pm 0.2$ [$\text{km}^{3}$], $n=3$ and $a=0.9$.  \label{fig2}}
\end{figure}

So, to better understand the nature of the possible real positive roots that (\ref{polyrot}) has, we must separate the cases as before. Nevertheless, as the polynomial can be factor out by $r^2$ the analysis is exactly the same given above for those cases where $l<0$ and $l>0$ for even and odd $n$. In Fig. \ref{fig2} the trend of the function $\Delta(r)$ against the radial coordinate is displayed. The upper panel shows the  solutions taking into account $l>0$, each one exhibiting their Cauchy horizon and the event horizon. On the other hand, the lower panel is depicting the case for the ``hairy'' BH when $l<0$. 

\begin{figure*}
\centering
\includegraphics[width=0.3\textwidth]{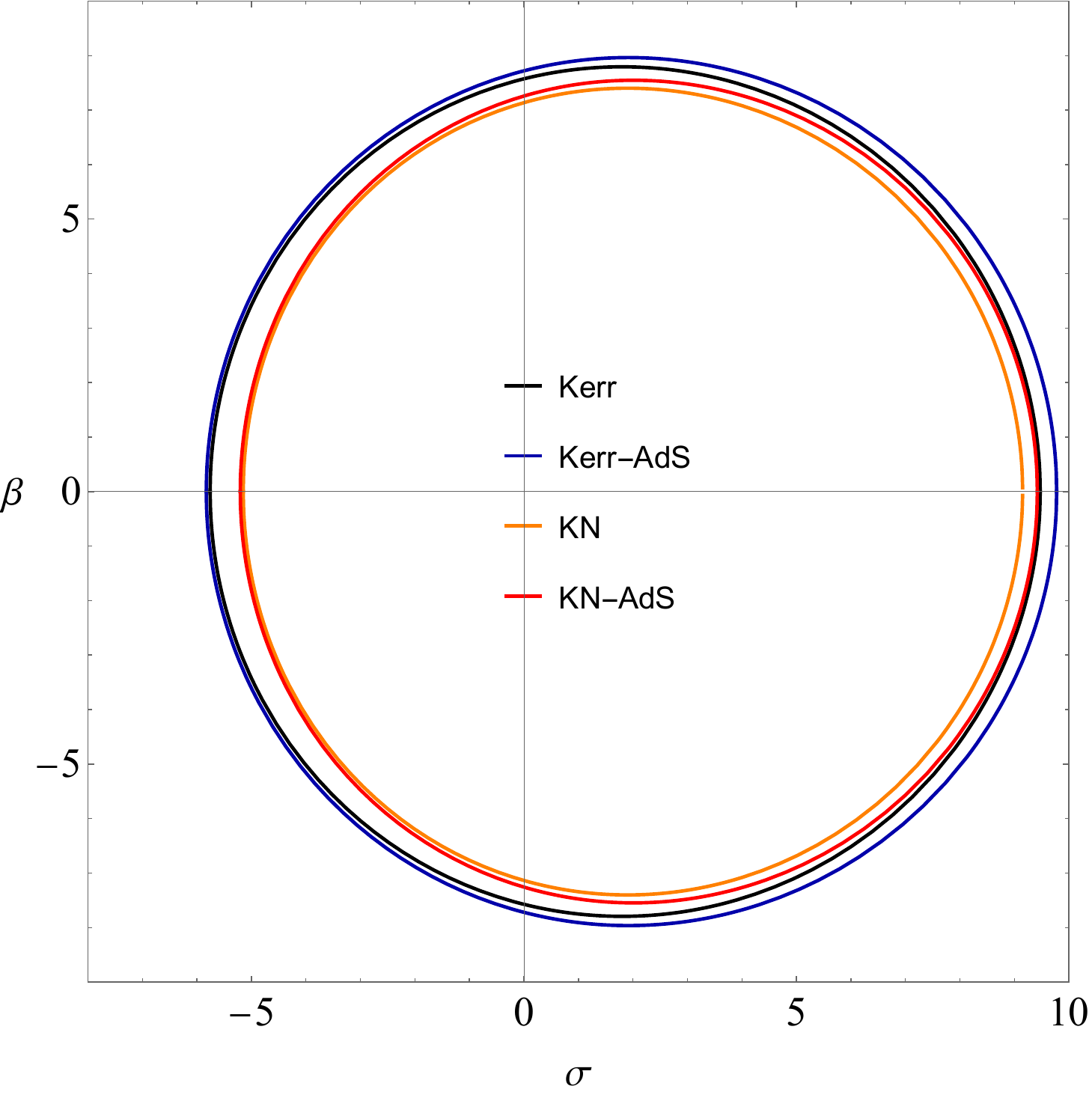} \
\includegraphics[width=0.3\textwidth]{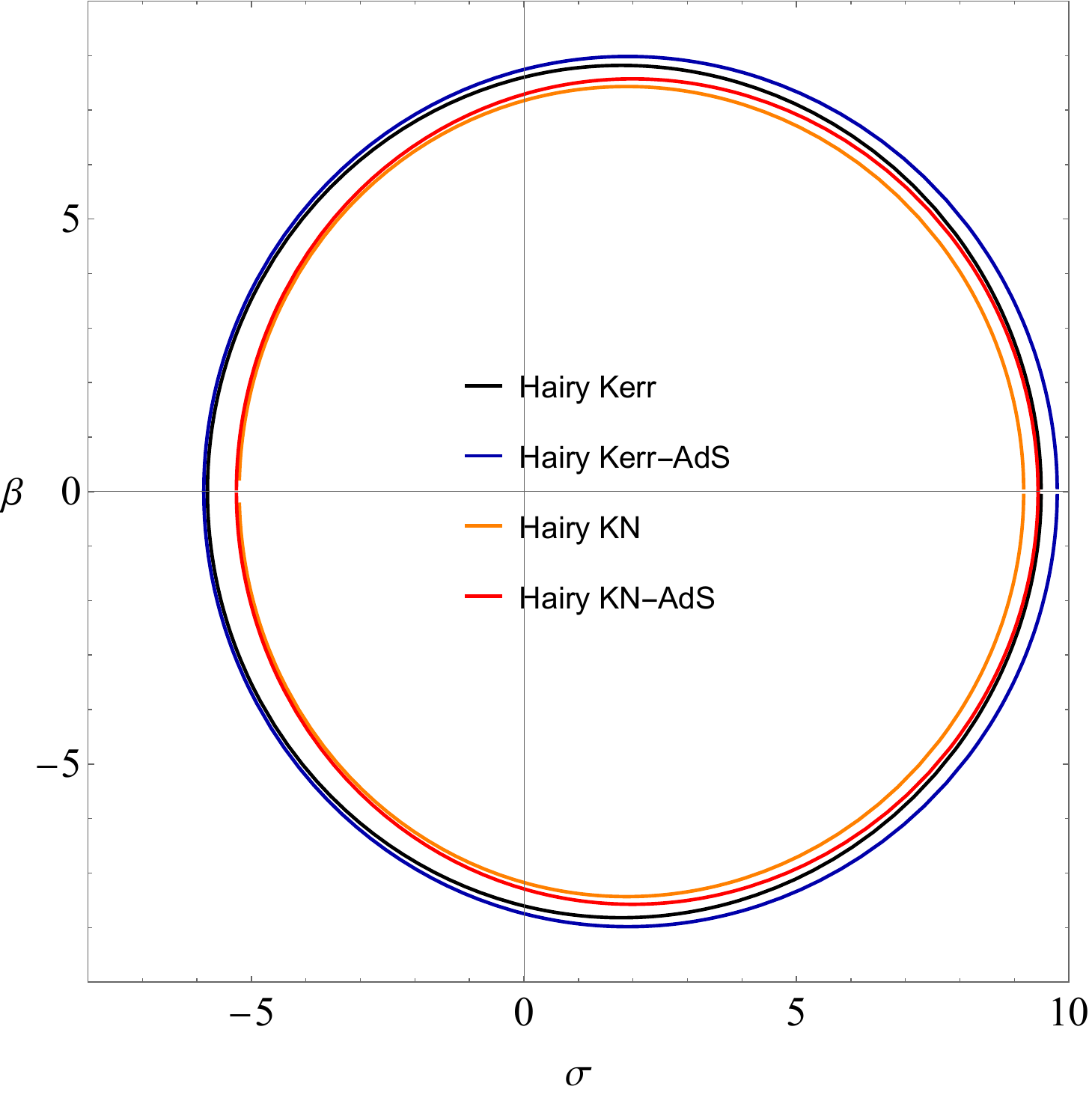} \ 
\includegraphics[width=0.3\textwidth]{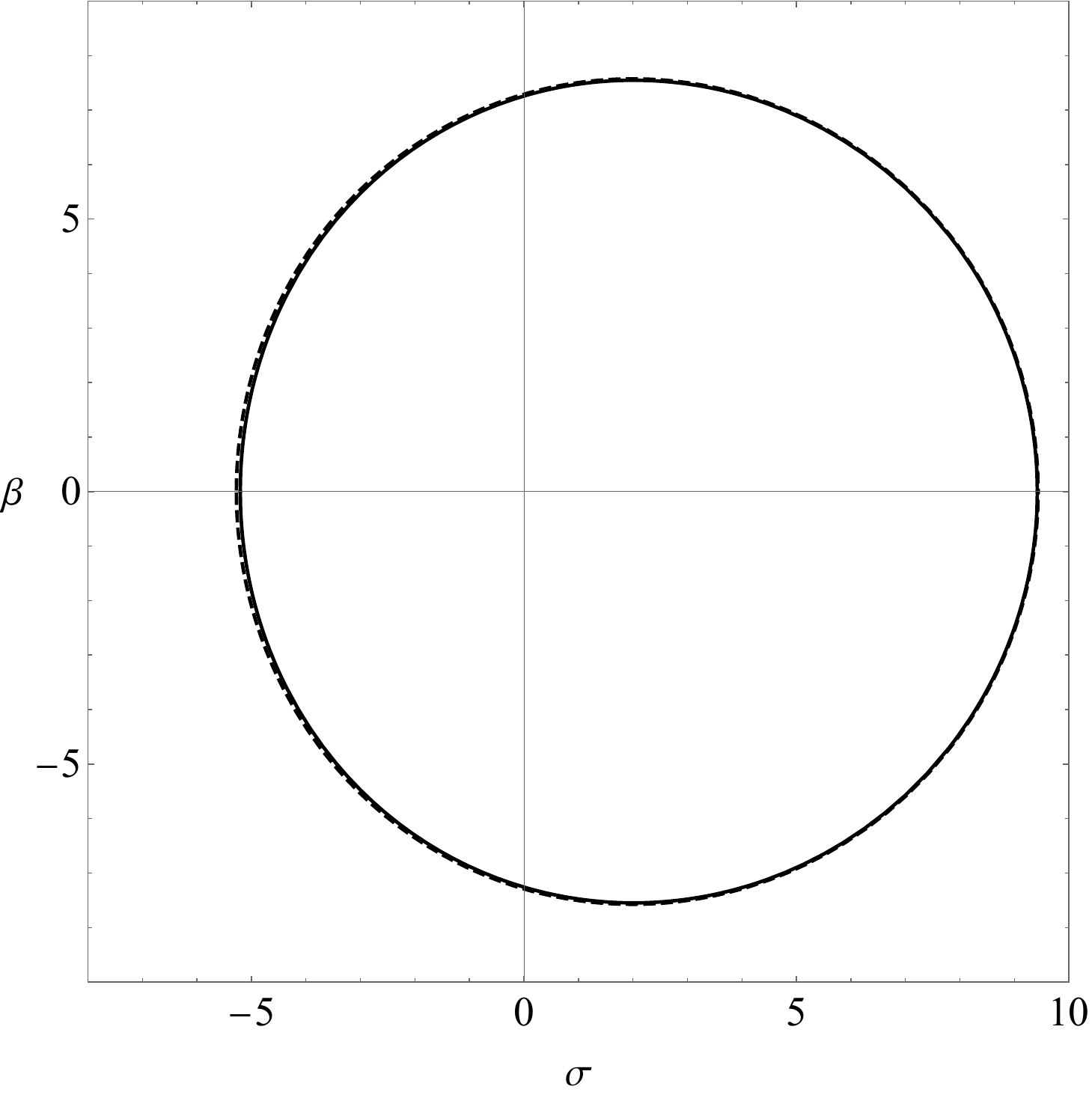}
\caption{ \textbf{Left panel}: The shadow of the seed space-times, Kerr, Kerr-dS, KN and KN-dS BHs. \textbf{Middle panel}: The shadow of the ``hairy'' space-times, Kerr, Kerr-dS, KN and KN-dS BHs. \textbf{Right panel}: A shadow comparison between KN-dS (solid line) and ``hairy'' KN-dS (dashed line) BHs. For all these plots we used $\mathcal{M}=1.5$ [km], $Q=0.8$ [km], $\Lambda=0.002$ [$\text{km}^{-2}$], $l=-0.2$ [$\text{km}^{3}$], $n=3$ and $a=0.9$. Here $\beta$ (vertical axis) and $\sigma$ (horizontal axis) are the celestial coordinates \cite{Vazquez:2003zm}. \label{fig3}}
\end{figure*}

Here there is only an event horizon. In comparison with the case $l=0$ it is evident that the horizon shifts
to larger radii when $l$ is present. This fact can influence the silhouette of the BH shadow. This is presented in Fig. \ref{fig3}, where the left panel shows the shadow of the Kerr, Kerr-dS, KN and KN-dS space-times. On the other hand, the central panel is exhibiting the shadow for the ``hairy'' versions of these BHs. As can be observed, they are following the same pattern, namely the ``hairy'' KN-dS rotates faster than the ``hairy'' Kerr BH\footnote{Notice that in comparison with the Kerr-dS and KN-dS there is a missing term $-\Lambda a^{2}r^{2}/3$ in the function $\Delta(r)$ (\ref{plo}). Then when $l=0$ one is not recovering the Kerr-dS and KN-dS, instead one gets other rotating BHs. Despite this fact, we are going to call these solutions as ``hairy'' Kerr and ``hairy'' KN-dS BHs, respectively.}. However, when comparing both the seed solution and the ``hairy'' solution, since the event horizon of the decoupled solutions is larger than its seed counterpart, then the shadow of the seed space-time is being screened by the shadow of the decoupled model (see right panel in Fig. \ref{fig3}). This means that the seed model rotates slightly faster than the deformed solution for the selected parameter space. This clearly is an effect of the presence of the parameter $l$ in the new solutions. Nevertheless, there is not a substantial different between the seminal and decoupled solution. 

\begin{figure}[H]
\centering
\includegraphics[width=0.45\textwidth]{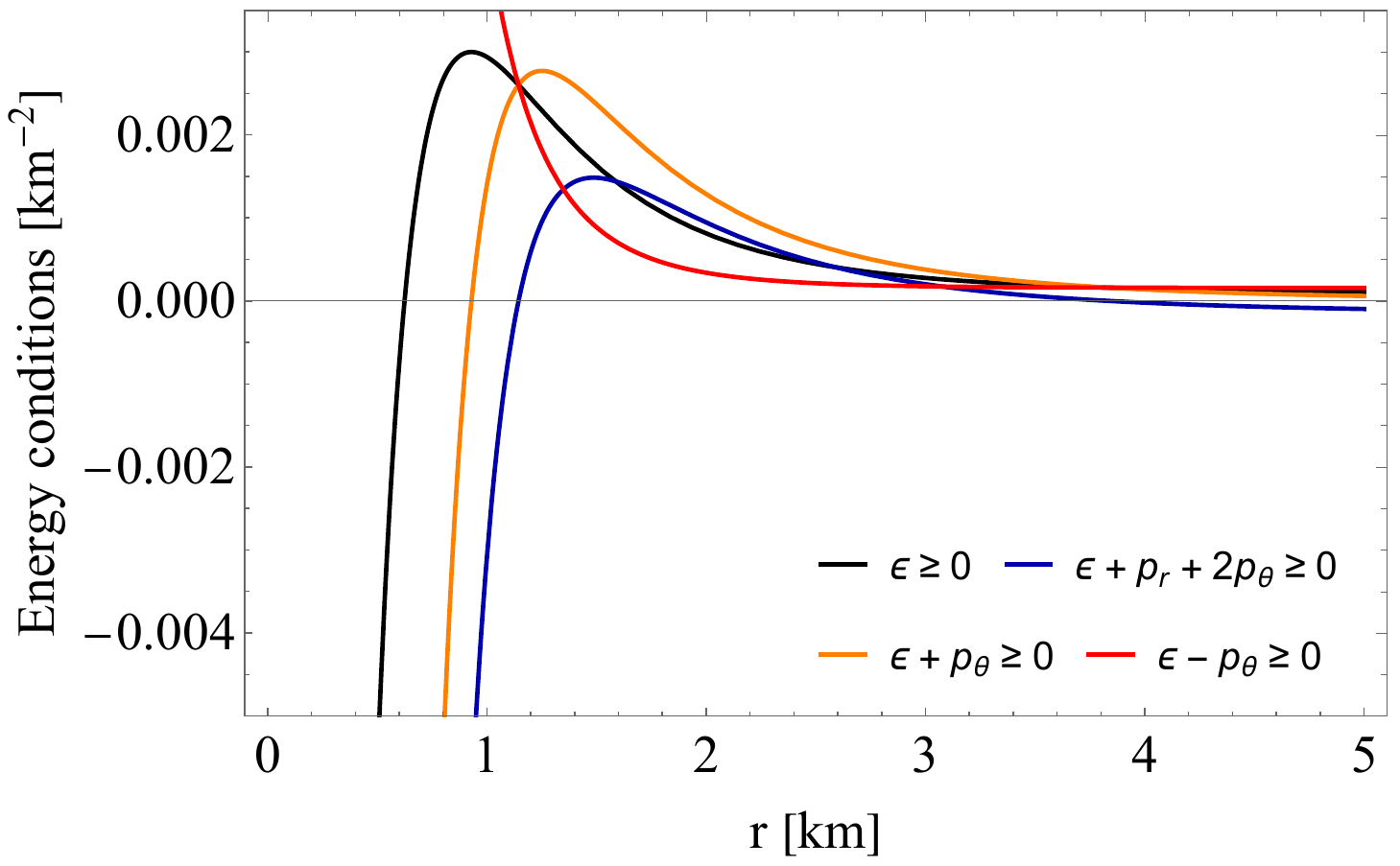}
\caption{ The trend of some inequalities corresponding to the strong and dominant energy conditions against the radial coordinate. For this plot we used $\mathcal{M}=1.5$ [km], $Q=0.8$ [km], $\Lambda=0.002$ [$\text{km}^{-2}$], $l=-0.2$ [$\text{km}^{3}$], $n=3$ and $a=0.9$. \label{fig6}}
\end{figure}

Now, we constraint the space parameter using the energy conditions as before. To get more insight about it, first of all one needs to compute the components of the energy-momentum tensor. In this regard, it is convenient to introduce the tetrads \cite{Ovalle:2021jzf,Contreras:2021yxe} 

\begin{equation}
\begin{aligned}
& {e}_t^\mu=\frac{\left(r^2+{a}^2, 0,0, {a}\right)}{\sqrt{\varrho^2 \Delta}}, \quad {e}_r^\mu=\frac{\sqrt{\Delta}(0,1,0,0)}{\sqrt{\varrho^2}} \\
& {e}_\theta^\mu=\frac{(0,0,1,0)}{\sqrt{\varrho^2}}, \quad {e}_\phi^\mu=-\frac{\left({a} \sin ^2 \theta, 0,0,1\right)}{\sqrt{\varrho^2} \sin \theta},
\end{aligned}
\end{equation}
leading to the following energy momentum-tensor, generating the metric (\ref{gg}),  

\begin{equation}
{T}^{\mu \nu}={\epsilon}(r) {e}_t^\mu {e}_t^\nu+{p}_r(r) {e}_r^\mu {e}_r^\nu+{p}_\theta(r) {e}_\theta^\mu {e}_\theta^\nu+{p}_\phi(r) {e}_\phi^\mu {e}_\phi^\nu,
\end{equation}
where the energy density $\epsilon(r)$ and the pressures $p_{r}(r)$, $p_{\theta}(r)$ and $p_{\phi}(r)$ are given by
\begin{equation}
\begin{aligned}\label{eq50}
{\epsilon}(r) & =-{p}_r(r)=\frac{2 r^2}{\kappa\varrho^4} {m}^{\prime}(r), \\
{p}_\theta(r) & ={p}_\phi(r)=-\frac{r}{\kappa\varrho^2} {m}^{\prime \prime}(r)+\frac{2\left(r^2-\varrho^2\right)}{\kappa\varrho^4} {m}^{\prime}(r).
\end{aligned}
\end{equation}
In this case, violation of SEC to prevent the formation of inner horizons, demands 
\begin{equation}\label{const1}
    p_{\theta}(r)<0\Rightarrow l<\frac{2\tilde{r}^{\, n}\left[\tilde{r}^{2}\left(2a^{2}+\tilde{r}^{2}\right)\Lambda-Q^{2}\right]}{\left[n-1\right]\left[a^{2}\left(n-2\right)+n\tilde{r}^{2}\right]},
\end{equation}
where in the limit $a=0$, the above constraint reduces to (\ref{const}). Besides, in this case the satisfaction  of (\ref{const1}) is subject to $0<\tilde{r}<a$. As it is shown in Fig. \ref{fig6}, the model has a positive energy density as expected due to $\Lambda>0$, what is more on of the components of the SEC (blue line) is violated in the BH region and all energy conditions are satisfied at the BH event horizon and beyond, except the SEC which is violated away the BH event horizon. This fact is typical for those space-times where $\Lambda>0$.

\section{Conclusions}\label{section5}

The Cauchy horizon is an inherent piece to those BHs characterized by electromagnetic charge or angular momentum (or both). This surface is pathological, and in most cases it is desirable to avoid it. In this regard, some effort has been put in to understand its consequences \cite{Dafermos:2003vim,German:2001ti} and ultimately prevent its formation \cite{Mars:2018hag,Mars:2018lup,Yang:2021civ,Cai:2020wrp,Devecioglu:2021xug,An:2021plu,Hartnoll:2020rwq,Carballo-Rubio:2018pmi}. In this article, from a pure classical treatment we avoided the formation of Cauchy horizons by introducing an unknown matter field. This was done via GD by e-MGD, where the Kerr-Schild ansatz and a general EoS were imposed to close the $\theta$-sector. This leads to a Schwarzschild-like space-time supplemented with a polynomial correction of the form $1/r^{n}$. Although these kinds of terms also appear in different contexts such as gravitational theories including high-order derivative terms \cite{Cano:2019ycn}, higher dimensions \cite{Agurto-Sepulveda:2022vvf}, $f(R)$ gravity \cite{Agrawal:2024wwt} or quantum gravity \cite{Lewandowski:2022zce}, the parameters accompanying these terms cannot be freely restricted. Then in most cases, BHs end up having at least one inner horizon. In the present case, the parameter $l$ can be considered as a ``secondary hair'' responsible for eliminating the Cauchy horizon in the case of RN, RN-dS, Kerr and KN solutions, leading to hairy BHs.
With respect to the so-called hairy Kerr-dS and KN-dS, these ones do not have a Cauchy horizon because of the presence of the parameter $l$ too. However, as stated, these solutions do not correspond to a genuine hairy extension of the original Kerr-dS and KN-dS. This is so because in the function $\Delta(r)$ there is a missing term $-\Lambda a^{2}r^{2}/3$ appearing in the seminal Kerr-dS and KN-dS BHs.  
On the other hand, for the Schwarzschild and Schwarzschild-dS no extra structure is added. This is a very important fact, since for all BHs containing a Cauchy horizon where the singularity is time-like, it becomes a space-like one and for those ones where their singularity is space-like in nature, this new term keeps this feature.     

With respect to some of the properties analyzed here for the rotating case, the presence of the new piece slows down the rotation of the BHs. This is so because the parameter $l$ shifts the event horizon to large values in comparison with the nondeformed case. To further support the parameter space considered here to avoid the existence of Cauchy horizons, we studied the energy conditions at the event horizon and its neighborhood. It is found that the dominant energy condition can be completely satisfied through the whole space-time, consequently the null and weak energy conditions.

Although it is a preliminary study, this holds promise in refining our understanding about hairy BHs with a simple causal structure and how the Cauchy horizon can be deleted using classical arguments. 
Clearly a more detailed study on optical, thermodynamic and stability properties is necessary to support the feasibility of the obtained solutions.

{Finally, it is important to mention the potential applicability of this methodology to obtain regular BHs without inner horizons. Although new solutions of regular BHs with a dS core have been reported using GD \cite{Ovalle:2023ref}, these toy models still have a Cauchy horizon. Of course, this requires a more thorough analysis, since in the case presented here, the correction that eliminates or avoids the appearance of inner horizons is clearly a nonregular term in the central zone of the BH. An unpleasant situation in the case of regular BH. All points and questions raised from this preliminary study will be investigated elsewhere.}


\bibliography{biblio.bib}

\begin{thebibliography}{10}
\expandafter\ifx\csname url\endcsname\relax
  \def\url#1{\texttt{#1}}\fi
\expandafter\ifx\csname urlprefix\endcsname\relax\def\urlprefix{URL }\fi
\expandafter\ifx\csname href\endcsname\relax
  \def\href#1#2{#2} \def\path#1{#1}\fi

\bibitem{EventHorizonTelescope:2019dse}
K.~Akiyama, et~al., {First M87 Event Horizon Telescope Results. I. The Shadow of the Supermassive Black Hole}, Astrophys. J. Lett. 875 (2019) L1.
\newblock \href {http://arxiv.org/abs/1906.11238} {\path{arXiv:1906.11238}}, \href {https://doi.org/10.3847/2041-8213/ab0ec7} {\path{doi:10.3847/2041-8213/ab0ec7}}.

\bibitem{EventHorizonTelescope:2019uob}
K.~Akiyama, et~al., {First M87 Event Horizon Telescope Results. II. Array and Instrumentation}, Astrophys. J. Lett. 875~(1) (2019) L2.
\newblock \href {http://arxiv.org/abs/1906.11239} {\path{arXiv:1906.11239}}, \href {https://doi.org/10.3847/2041-8213/ab0c96} {\path{doi:10.3847/2041-8213/ab0c96}}.

\bibitem{EventHorizonTelescope:2019jan}
K.~Akiyama, et~al., {First M87 Event Horizon Telescope Results. III. Data Processing and Calibration}, Astrophys. J. Lett. 875~(1) (2019) L3.
\newblock \href {http://arxiv.org/abs/1906.11240} {\path{arXiv:1906.11240}}, \href {https://doi.org/10.3847/2041-8213/ab0c57} {\path{doi:10.3847/2041-8213/ab0c57}}.

\bibitem{EventHorizonTelescope:2019ths}
K.~Akiyama, et~al., {First M87 Event Horizon Telescope Results. IV. Imaging the Central Supermassive Black Hole}, Astrophys. J. Lett. 875~(1) (2019) L4.
\newblock \href {http://arxiv.org/abs/1906.11241} {\path{arXiv:1906.11241}}, \href {https://doi.org/10.3847/2041-8213/ab0e85} {\path{doi:10.3847/2041-8213/ab0e85}}.

\bibitem{EventHorizonTelescope:2019pgp}
K.~Akiyama, et~al., {First M87 Event Horizon Telescope Results. V. Physical Origin of the Asymmetric Ring}, Astrophys. J. Lett. 875~(1) (2019) L5.
\newblock \href {http://arxiv.org/abs/1906.11242} {\path{arXiv:1906.11242}}, \href {https://doi.org/10.3847/2041-8213/ab0f43} {\path{doi:10.3847/2041-8213/ab0f43}}.

\bibitem{EventHorizonTelescope:2019ggy}
K.~Akiyama, et~al., {First M87 Event Horizon Telescope Results. VI. The Shadow and Mass of the Central Black Hole}, Astrophys. J. Lett. 875~(1) (2019) L6.
\newblock \href {http://arxiv.org/abs/1906.11243} {\path{arXiv:1906.11243}}, \href {https://doi.org/10.3847/2041-8213/ab1141} {\path{doi:10.3847/2041-8213/ab1141}}.

\bibitem{Vagnozzi:2022moj}
S.~Vagnozzi, et~al., {Horizon-scale tests of gravity theories and fundamental physics from the Event Horizon Telescope image of Sagittarius A}, Class. Quant. Grav. 40~(16) (2023) 165007.
\newblock \href {http://arxiv.org/abs/2205.07787} {\path{arXiv:2205.07787}}, \href {https://doi.org/10.1088/1361-6382/acd97b} {\path{doi:10.1088/1361-6382/acd97b}}.

\bibitem{Hawking:1971vc}
S.~W. Hawking, {Black holes in general relativity}, Commun. Math. Phys. 25 (1972) 152--166.
\newblock \href {https://doi.org/10.1007/BF01877517} {\path{doi:10.1007/BF01877517}}.

\bibitem{Ruffini:1971bza}
R.~Ruffini, J.~A. Wheeler, {Introducing the black hole}, Phys. Today 24~(1) (1971) 30.
\newblock \href {https://doi.org/10.1063/1.3022513} {\path{doi:10.1063/1.3022513}}.

\bibitem{Hawking:2016msc}
S.~W. Hawking, M.~J. Perry, A.~Strominger, {Soft Hair on Black Holes}, Phys. Rev. Lett. 116~(23) (2016) 231301.
\newblock \href {http://arxiv.org/abs/1601.00921} {\path{arXiv:1601.00921}}, \href {https://doi.org/10.1103/PhysRevLett.116.231301} {\path{doi:10.1103/PhysRevLett.116.231301}}.

\bibitem{Dadhich:2020ukj}
N.~Dadhich, {On causal structure of $4D$-Einstein\textendash{}Gauss\textendash{}Bonnet black hole}, Eur. Phys. J. C 80~(9) (2020) 832.
\newblock \href {http://arxiv.org/abs/2005.05757} {\path{arXiv:2005.05757}}, \href {https://doi.org/10.1140/epjc/s10052-020-8422-8} {\path{doi:10.1140/epjc/s10052-020-8422-8}}.

\bibitem{Penrose:1964wq}
R.~Penrose, {Gravitational collapse and space-time singularities}, Phys. Rev. Lett. 14 (1965) 57--59.
\newblock \href {https://doi.org/10.1103/PhysRevLett.14.57} {\path{doi:10.1103/PhysRevLett.14.57}}.

\bibitem{Penrose:1969pc}
R.~Penrose, {Gravitational collapse: The role of general relativity}, Riv. Nuovo Cim. 1 (1969) 252--276.
\newblock \href {https://doi.org/10.1023/A:1016578408204} {\path{doi:10.1023/A:1016578408204}}.

\bibitem{Chesler:2019tco}
P.~M. Chesler, R.~Narayan, E.~Curiel, {Singularities in Reissner\textendash{}Nordstr\"om black holes}, Class. Quant. Grav. 37~(2) (2020) 025009.
\newblock \href {http://arxiv.org/abs/1902.08323} {\path{arXiv:1902.08323}}, \href {https://doi.org/10.1088/1361-6382/ab5b69} {\path{doi:10.1088/1361-6382/ab5b69}}.

\bibitem{Davey:2024xvd}
A.~Davey, O.~J.~C. Dias, D.~S. Gil, {Strong Cosmic Censorship in Kerr-Newman-de Sitter}, JHEP 07 (2024) 113.
\newblock \href {http://arxiv.org/abs/2404.03724} {\path{arXiv:2404.03724}}, \href {https://doi.org/10.1007/JHEP07(2024)113} {\path{doi:10.1007/JHEP07(2024)113}}.

\bibitem{Ong:2020xwv}
Y.~C. Ong, {Space\textendash{}time singularities and cosmic censorship conjecture: A Review with some thoughts}, Int. J. Mod. Phys. A 35~(14) (2020) 14.
\newblock \href {http://arxiv.org/abs/2005.07032} {\path{arXiv:2005.07032}}, \href {https://doi.org/10.1142/S0217751X20300070} {\path{doi:10.1142/S0217751X20300070}}.

\bibitem{Ovalle:2017fgl}
J.~Ovalle, {Decoupling gravitational sources in general relativity: from perfect to anisotropic fluids}, Phys. Rev. D 95~(10) (2017) 104019.
\newblock \href {https://doi.org/10.1103/PhysRevD.95.104019} {\path{doi:10.1103/PhysRevD.95.104019}}.

\bibitem{Ovalle:2018gic}
J.~Ovalle, {Decoupling gravitational sources in general relativity: The extended case}, Phys. Lett. B 788 (2019) 213--218.
\newblock \href {http://arxiv.org/abs/1812.03000} {\path{arXiv:1812.03000}}, \href {https://doi.org/10.1016/j.physletb.2018.11.029} {\path{doi:10.1016/j.physletb.2018.11.029}}.

\bibitem{Ovalle:2020fuo}
J.~Ovalle, R.~Casadio, {Beyond Einstein Gravity}: {The Minimal Geometric Deformation Approach in the Brane-World}, SpringerBriefs in Physics, Springer, 2020.
\newblock \href {https://doi.org/10.1007/978-3-030-39493-6} {\path{doi:10.1007/978-3-030-39493-6}}.

\bibitem{Heydarzade:2023dof}
Y.~Heydarzade, M.~Misyura, V.~Vertogradov, {Hairy Kiselev black hole solutions}, Phys. Rev. D 108~(4) (2023) 044073.
\newblock \href {http://arxiv.org/abs/2307.04556} {\path{arXiv:2307.04556}}, \href {https://doi.org/10.1103/PhysRevD.108.044073} {\path{doi:10.1103/PhysRevD.108.044073}}.

\bibitem{Ovalle:2018umz}
J.~Ovalle, R.~Casadio, R.~d. Rocha, A.~Sotomayor, Z.~Stuchlik, {Black holes by gravitational decoupling}, Eur. Phys. J. C 78~(11) (2018) 960.
\newblock \href {http://arxiv.org/abs/1804.03468} {\path{arXiv:1804.03468}}, \href {https://doi.org/10.1140/epjc/s10052-018-6450-4} {\path{doi:10.1140/epjc/s10052-018-6450-4}}.

\bibitem{Cavalcanti:2022adb}
R.~T. Cavalcanti, K.~d.~S. Alves, J.~M. Hoff~da Silva, {Near-Horizon Thermodynamics of Hairy Black Holes from Gravitational Decoupling}, Universe 8~(7) (2022) 363.
\newblock \href {http://arxiv.org/abs/2207.03995} {\path{arXiv:2207.03995}}, \href {https://doi.org/10.3390/universe8070363} {\path{doi:10.3390/universe8070363}}.

\bibitem{Cavalcanti:2022cga}
R.~T. Cavalcanti, R.~C. de~Paiva, R.~da~Rocha, {Echoes of the gravitational decoupling: scalar perturbations and quasinormal modes of hairy black holes}, Eur. Phys. J. Plus 137~(10) (2022) 1185.
\newblock \href {http://arxiv.org/abs/2203.08740} {\path{arXiv:2203.08740}}, \href {https://doi.org/10.1140/epjp/s13360-022-03407-x} {\path{doi:10.1140/epjp/s13360-022-03407-x}}.

\bibitem{Meert:2021khi}
P.~Meert, R.~da~Rocha, {Gravitational decoupling, hairy black holes and conformal anomalies}, Eur. Phys. J. C 82~(2) (2022) 175.
\newblock \href {http://arxiv.org/abs/2109.06289} {\path{arXiv:2109.06289}}, \href {https://doi.org/10.1140/epjc/s10052-022-10121-6} {\path{doi:10.1140/epjc/s10052-022-10121-6}}.

\bibitem{Sultana:2021cvq}
J.~Sultana, {Gravitational Decoupling in Higher Order Theories}, Symmetry 13~(9) (2021) 1598.
\newblock \href {https://doi.org/10.3390/sym13091598} {\path{doi:10.3390/sym13091598}}.

\bibitem{Ovalle:2020kpd}
J.~Ovalle, R.~Casadio, E.~Contreras, A.~Sotomayor, {Hairy black holes by gravitational decoupling}, Phys. Dark Univ. 31 (2021) 100744.
\newblock \href {http://arxiv.org/abs/2006.06735} {\path{arXiv:2006.06735}}, \href {https://doi.org/10.1016/j.dark.2020.100744} {\path{doi:10.1016/j.dark.2020.100744}}.

\bibitem{daRocha:2020gee}
R.~a. da~Rocha, A.~A. Tomaz, {MGD-decoupled black holes, anisotropic fluids and holographic entanglement entropy}, Eur. Phys. J. C 80~(9) (2020) 857.
\newblock \href {http://arxiv.org/abs/2005.02980} {\path{arXiv:2005.02980}}, \href {https://doi.org/10.1140/epjc/s10052-020-8414-8} {\path{doi:10.1140/epjc/s10052-020-8414-8}}.

\bibitem{Fernandes-Silva:2019fez}
A.~Fernandes-Silva, A.~J. Ferreira-Martins, R.~da~Rocha, {Extended quantum portrait of MGD black holes and information entropy}, Phys. Lett. B 791 (2019) 323--330.
\newblock \href {http://arxiv.org/abs/1901.07492} {\path{arXiv:1901.07492}}, \href {https://doi.org/10.1016/j.physletb.2019.03.010} {\path{doi:10.1016/j.physletb.2019.03.010}}.

\bibitem{Casadio:2022ndh}
R.~Casadio, A.~Giusti, J.~Ovalle, {Quantum Reissner-Nordstr\"om geometry: Singularity and Cauchy horizon}, Phys. Rev. D 105~(12) (2022) 124026.
\newblock \href {http://arxiv.org/abs/2203.03252} {\path{arXiv:2203.03252}}, \href {https://doi.org/10.1103/PhysRevD.105.124026} {\path{doi:10.1103/PhysRevD.105.124026}}.

\bibitem{Ovalle:2022eqb}
J.~Ovalle, {Warped vacuum energy by black holes}, Eur. Phys. J. C 82~(2) (2022) 170.
\newblock \href {http://arxiv.org/abs/2202.12037} {\path{arXiv:2202.12037}}, \href {https://doi.org/10.1140/epjc/s10052-022-10094-6} {\path{doi:10.1140/epjc/s10052-022-10094-6}}.

\bibitem{Estrada:2020ptc}
M.~Estrada, R.~Prado, {A note of the first law of thermodynamics by gravitational decoupling}, Eur. Phys. J. C 80~(8) (2020) 799.
\newblock \href {http://arxiv.org/abs/2003.13168} {\path{arXiv:2003.13168}}, \href {https://doi.org/10.1140/epjc/s10052-020-8315-x} {\path{doi:10.1140/epjc/s10052-020-8315-x}}.

\bibitem{Estrada:2021kuj}
M.~Estrada, {Gravitational Decoupling algorithm modifies the value of the conserved charges and thermodynamics properties in Lovelock Unique Vacuum theory}, Annals Phys. 439 (2022) 168792.
\newblock \href {http://arxiv.org/abs/2106.02166} {\path{arXiv:2106.02166}}, \href {https://doi.org/10.1016/j.aop.2022.168792} {\path{doi:10.1016/j.aop.2022.168792}}.

\bibitem{Ovalle:2023ref}
J.~Ovalle, R.~Casadio, A.~Giusti, {Regular hairy black holes through Minkowski deformation}, Phys. Lett. B 844 (2023) 138085.
\newblock \href {http://arxiv.org/abs/2304.03263} {\path{arXiv:2304.03263}}, \href {https://doi.org/10.1016/j.physletb.2023.138085} {\path{doi:10.1016/j.physletb.2023.138085}}.

\bibitem{Zhang:2022niv}
C.-M. Zhang, M.~Zhang, D.-C. Zou, {Gravitational decoupling for hairy black holes in asymptotic AdS spacetimes*}, Chin. Phys. C 47~(1) (2023) 015106.
\newblock \href {http://arxiv.org/abs/2208.06830} {\path{arXiv:2208.06830}}, \href {https://doi.org/10.1088/1674-1137/ac9b2c} {\path{doi:10.1088/1674-1137/ac9b2c}}.

\bibitem{Khosravipoor:2023jsl}
M.~R. Khosravipoor, M.~Farhoudi, {Thermodynamics of deformed AdS-Schwarzschild black hole}, Eur. Phys. J. C 83~(11) (2023) 1045.
\newblock \href {http://arxiv.org/abs/2311.02456} {\path{arXiv:2311.02456}}, \href {https://doi.org/10.1140/epjc/s10052-023-12222-2} {\path{doi:10.1140/epjc/s10052-023-12222-2}}.

\bibitem{Casadio:2024uwj}
R.~Casadio, C.~N. Souza, R.~da~Rocha, {Gravitational decoupling and aerodynamics: black holes and analog gravity in a jet propulsion lab}, Eur. Phys. J. C 84~(8) (2024) 767.
\newblock \href {http://arxiv.org/abs/2402.04682} {\path{arXiv:2402.04682}}, \href {https://doi.org/10.1140/epjc/s10052-024-13131-8} {\path{doi:10.1140/epjc/s10052-024-13131-8}}.

\bibitem{Liang:2024xif}
Y.~Liang, X.~Lyu, J.~Tao, {Observational appearances of hairy black holes in the framework of gravitational decoupling}, Commun. Theor. Phys. 76~(8) (2024) 085402.
\newblock \href {https://doi.org/10.1088/1572-9494/ad4ce0} {\path{doi:10.1088/1572-9494/ad4ce0}}.

\bibitem{Ovalle:2021jzf}
J.~Ovalle, E.~Contreras, Z.~Stuchlik, {Kerr\textendash{}de Sitter black hole revisited}, Phys. Rev. D 103~(8) (2021) 084016.
\newblock \href {http://arxiv.org/abs/2104.06359} {\path{arXiv:2104.06359}}, \href {https://doi.org/10.1103/PhysRevD.103.084016} {\path{doi:10.1103/PhysRevD.103.084016}}.

\bibitem{Contreras:2021yxe}
E.~Contreras, J.~Ovalle, R.~Casadio, {Gravitational decoupling for axially symmetric systems and rotating black holes}, Phys. Rev. D 103~(4) (2021) 044020.
\newblock \href {http://arxiv.org/abs/2101.08569} {\path{arXiv:2101.08569}}, \href {https://doi.org/10.1103/PhysRevD.103.044020} {\path{doi:10.1103/PhysRevD.103.044020}}.

\bibitem{Ovalle:2022yjl}
J.~Ovalle, E.~Contreras, Z.~Stuchlik, {Energy exchange between relativistic fluids: the polytropic case}, Eur. Phys. J. C 82~(3) (2022) 211.
\newblock \href {http://arxiv.org/abs/2202.12665} {\path{arXiv:2202.12665}}, \href {https://doi.org/10.1140/epjc/s10052-022-10168-5} {\path{doi:10.1140/epjc/s10052-022-10168-5}}.

\bibitem{Kerr:1965wfc}
R.~P. Kerr, A.~Schild, {Some algebraically degenerate solutions of Einstein\textquoteright{}s gravitational field equations}, Proc. Symp. Appl. Math. 17 (1965) 199.

\bibitem{Yang:2021civ}
R.-Q. Yang, R.-G. Cai, L.~Li, {Constraining the number of horizons with energy conditions}, Class. Quant. Grav. 39~(3) (2022) 035005.
\newblock \href {http://arxiv.org/abs/2104.03012} {\path{arXiv:2104.03012}}, \href {https://doi.org/10.1088/1361-6382/ac4118} {\path{doi:10.1088/1361-6382/ac4118}}.

\bibitem{Curiel:2014zba}
E.~Curiel, {A Primer on Energy Conditions}, Einstein Stud. 13 (2017) 43--104.
\newblock \href {https://doi.org/10.1007/978-1-4939-3210-8_3} {\path{doi:10.1007/978-1-4939-3210-8_3}}.

\bibitem{Vazquez:2003zm}
S.~E. Vazquez, E.~P. Esteban, {Strong field gravitational lensing by a Kerr black hole}, Nuovo Cim. B 119 (2004) 489--519.
\newblock \href {http://arxiv.org/abs/gr-qc/0308023} {\path{arXiv:gr-qc/0308023}}, \href {https://doi.org/10.1393/ncb/i2004-10121-y} {\path{doi:10.1393/ncb/i2004-10121-y}}.

\bibitem{Dafermos:2003vim}
M.~Dafermos, {Stability and Instability of the Cauchy Horizon for the Spherically Symmetric Einstein-Maxwell-Scalar Field Equations}, Ann. Math 158~(3) (2003) 875--928.

\bibitem{German:2001ti}
W.~S. German, I.~G. Moss, {Cauchy horizon stability and cosmic censorship}, Class. Quant. Grav. 18 (2001) 5097--5102.
\newblock \href {http://arxiv.org/abs/gr-qc/0103080} {\path{arXiv:gr-qc/0103080}}, \href {https://doi.org/10.1088/0264-9381/18/23/306} {\path{doi:10.1088/0264-9381/18/23/306}}.

\bibitem{Mars:2018hag}
M.~Mars, T.-T. Paetz, J.~Senovilla, M.~M., {Multiple Killing Horizons}, Class. Quant. Grav. 35~(15) (2018) 155015.
\newblock \href {http://arxiv.org/abs/1803.03054} {\path{arXiv:1803.03054}}, \href {https://doi.org/10.1088/1361-6382/aacd2c} {\path{doi:10.1088/1361-6382/aacd2c}}.

\bibitem{Mars:2018lup}
M.~Mars, T.-T. Paetz, J.~M.~M. Senovilla, {Multiple Killing Horizons and Near Horizon Geometries}, Class. Quant. Grav. 35~(24) (2018) 245007.
\newblock \href {http://arxiv.org/abs/1807.02679} {\path{arXiv:1807.02679}}, \href {https://doi.org/10.1088/1361-6382/aaeaf1} {\path{doi:10.1088/1361-6382/aaeaf1}}.

\bibitem{Cai:2020wrp}
R.-G. Cai, L.~Li, R.-Q. Yang, {No Inner-Horizon Theorem for Black Holes with Charged Scalar Hairs}, JHEP 03 (2021) 263.
\newblock \href {http://arxiv.org/abs/2009.05520} {\path{arXiv:2009.05520}}, \href {https://doi.org/10.1007/JHEP03(2021)263} {\path{doi:10.1007/JHEP03(2021)263}}.

\bibitem{Devecioglu:2021xug}
D.~O. Devecioglu, M.-I. Park, {No scalar-haired Cauchy horizon theorem in Einstein-Maxwell-Horndeski theories}, Phys. Lett. B 829 (2022) 137107.
\newblock \href {http://arxiv.org/abs/2101.10116} {\path{arXiv:2101.10116}}, \href {https://doi.org/10.1016/j.physletb.2022.137107} {\path{doi:10.1016/j.physletb.2022.137107}}.

\bibitem{An:2021plu}
Y.-S. An, L.~Li, F.-G. Yang, {No Cauchy horizon theorem for nonlinear electrodynamics black holes with charged scalar hairs}, Phys. Rev. D 104~(2) (2021) 024040.
\newblock \href {http://arxiv.org/abs/2106.01069} {\path{arXiv:2106.01069}}, \href {https://doi.org/10.1103/PhysRevD.104.024040} {\path{doi:10.1103/PhysRevD.104.024040}}.

\bibitem{Hartnoll:2020rwq}
S.~A. Hartnoll, G.~T. Horowitz, J.~Kruthoff, J.~E. Santos, {Gravitational duals to the grand canonical ensemble abhor Cauchy horizons}, JHEP 10 (2020) 102.
\newblock \href {http://arxiv.org/abs/2006.10056} {\path{arXiv:2006.10056}}, \href {https://doi.org/10.1007/JHEP10(2020)102} {\path{doi:10.1007/JHEP10(2020)102}}.

\bibitem{Carballo-Rubio:2018pmi}
R.~Carballo-Rubio, F.~Di~Filippo, S.~Liberati, C.~Pacilio, M.~Visser, {On the viability of regular black holes}, JHEP 07 (2018) 023.
\newblock \href {http://arxiv.org/abs/1805.02675} {\path{arXiv:1805.02675}}, \href {https://doi.org/10.1007/JHEP07(2018)023} {\path{doi:10.1007/JHEP07(2018)023}}.

\bibitem{Cano:2019ycn}
P.~A. Cano, S.~Chimento, R.~Linares, T.~Ort\'\i{}n, P.~F. Ram\'\i{}rez, {$\alpha'$ corrections of Reissner-Nordstr\"om black holes}, JHEP 02 (2020) 031.
\newblock \href {http://arxiv.org/abs/1910.14324} {\path{arXiv:1910.14324}}, \href {https://doi.org/10.1007/JHEP02(2020)031} {\path{doi:10.1007/JHEP02(2020)031}}.

\bibitem{Agurto-Sepulveda:2022vvf}
F.~Agurto-Sep\'ulveda, M.~Chernicoff, G.~Giribet, J.~Oliva, M.~Oyarzo, {Slowly rotating and accelerating \ensuremath{\alpha}'-corrected black holes in four and higher dimensions}, Phys. Rev. D 107~(8) (2023) 084014.
\newblock \href {http://arxiv.org/abs/2207.13214} {\path{arXiv:2207.13214}}, \href {https://doi.org/10.1103/PhysRevD.107.084014} {\path{doi:10.1103/PhysRevD.107.084014}}.

\bibitem{Agrawal:2024wwt}
A.~S. Agrawal, S.~Zerbini, B.~Mishra, {Black holes and wormholes beyond classical general relativity}, Phys. Dark Univ. 46 (2024) 101637.
\newblock \href {http://arxiv.org/abs/2406.01241} {\path{arXiv:2406.01241}}, \href {https://doi.org/10.1016/j.dark.2024.101637} {\path{doi:10.1016/j.dark.2024.101637}}.

\bibitem{Lewandowski:2022zce}
J.~Lewandowski, Y.~Ma, J.~Yang, C.~Zhang, {Quantum Oppenheimer-Snyder and Swiss Cheese Models}, Phys. Rev. Lett. 130~(10) (2023) 101501.
\newblock \href {http://arxiv.org/abs/2210.02253} {\path{arXiv:2210.02253}}, \href {https://doi.org/10.1103/PhysRevLett.130.101501} {\path{doi:10.1103/PhysRevLett.130.101501}}.

\end{thebibliography}
\bibliographystyle{elsarticle-num}

\end{document}